\documentclass[11pt]{article}
\pdfoutput=1
\usepackage{jheppub}
\usepackage{amsmath}
\usepackage{bm}
\usepackage{slashed}
\usepackage{graphicx}
\usepackage{enumerate}

\usepackage{longtable}
\usepackage{pifont}
\usepackage{ytableau}

\newlength{\dummysp}
\newcommand{\beq}{\begin{eqnarray}}
\newcommand{\eeq}{\end{eqnarray}}

\newcommand{\gappeq}{\mathrel{\rlap {\raise.5ex\hbox{$>$}}
{\lower.5ex\hbox{$\sim$}}}}
\newcommand{\lappeq}{\mathrel{\rlap{\raise.5ex\hbox{$<$}}
{\lower.5ex\hbox{$\sim$}}}}

\newcommand{\ben}{\begin{enumerate}}
\newcommand{\een}{\end{enumerate}}

\newcommand{\bit}{\begin{itemize}}
\newcommand{\eit}{\end{itemize}}

\def\[{\left [}
\def\]{\right ]}
\def\({\left (}
\def\){\right )}

\title{Entanglement entropy, dualities, and deconfinement in gauge theories\\
 }
   
\author[a]{Mohamed M. Anber,}\author[a]{Benjamin J. Kolligs} 
  
\affiliation[a]{Department of Physics, Lewis \& Clark College, 
Portland, OR 97219, USA} 
\emailAdd{manber@lclark.edu}\emailAdd{benjaminkolligs@lclark.edu}  
  
\abstract{Computing the entanglement entropy in confining gauge theories is often accompanied by puzzles and ambiguities. In this work we show that compactifying the theory on a small circle $\mathbb S^1_L$ evades these difficulties. In particular, we study Yang-Mills theory on $\mathbb R^3\times \mathbb S^1_L$ with double-trace deformations or adjoint fermions and hold it at temperatures near the deconfinement transition. This theory is dual to a multi-component (electric-magnetic) Coulomb gas that can be mapped either to an XY-spin model with $\mathbb Z_p$ symmetry-preserving perturbations or dual Sine-Gordon model. The entanglement entropy of the dual Sine-Gordon model exhibits an extremum at the critical temperature/crossover. We also compute R\'enyi mutual information (RMI) of the XY-spin model by means of the replica trick and Monte Carlo simulations. These are expensive calculations, since one in general needs to suppress lower winding vortices that do not correspond to physical excitations of the system. We use a T-duality that maps the original XY model to its mirror image, making the extraction of RMI a much efficient process. Our simulations indicate that RMI follows the area law scaling, with subleading corrections, and this quantity can be used as a genuine probe to detect deconfinement transitions. We also discuss the effect of fundamental matter on RMI and the implications of our findings in gauge theories and beyond.}

\begin{document}

\maketitle

\flushbottom

%%%%%%%%%%%%%%%%%%%%%%%%%%%%%%%%%%%%%
\section{Introduction}
%%%%%%%%%%%%%%%%%%%%%%%%%%%%%%%%%%%%%

Information-theoretic techniques in quantum/statistical field theory has become an increasingly important tool for studying quantum as well as classical phases of matter \cite{Laflorencie:2015eck,2015arXiv150802595Z}. The power of information theory is that it provides probes that are able to distinguish between different phases, even in the absence of  local order parameters. This is attributed to the fact that the information encrypted in a system is independent of the nature of its fundamental constituents. 

In the simplest setup,  one uses correlation functions, ${C}(x,y)$, of fields that appear in a Lagrangian to form probes that transform non-trivially under certain global symmetries.  ${C}(x,y)$ tells us how different parts of the system correlate to eachother as the system transits from one phase to the other. In certain situations, however, there exists no global symmetries or that  ${C}(x,y)$ is not sufficient to characterize the correlation in the system. In these cases  quantities like entanglement entropy and mutual information are indispensable for studying various quantum and classical phase transitions. 

Continuum and lattice gauge theories have also been investigated in the light of information theory, with often puzzling conclusions \cite{Kabat:1995eq,Buividovich:2008kq,Zhitnitsky:2011tr,Solodukhin:2012jh,Donnelly:2012st,Eling:2013aqa,Agon:2013iva,Huang:2014pfa,Donnelly:2016mlc}. The complexity of gauge theories stems from the fact that they are invariant under gauge redundancies and one needs to be careful to account only for the physical rather than the spurious degrees of freedom. Moreover, nonabelian asymptotically free theories are strongly coupled in the IR, making the calculations of entanglement entropy a rather daunting task. A lattice formulation of the problem is also plagued with ambiguities, since the gauge invariant Hilbert space cannot be factorized into a tensor product of gauge invariant subspaces and one needs to extend the definition of the Hilbert space. This fact was first realized in \cite{Buividovich:2008kq} and then further investigated in subsequent works, see e.g., \cite{Casini:2013rba,Radicevic:2014kqa,Ghosh:2015iwa,Aoki:2015bsa,VanAcoleyen:2015ccp}. Such difficulties may be circumvented by invoking the gravity dual, as was first proposed in \cite{Ryu:2006bv,Nishioka:2006gr,Klebanov:2007ws} and further examined in many works, see, e.g., \cite{Fujita:2008zv,Buividovich:2008kq,Velytsky:2008rs}. In these works it was argued that  the entanglement entropy between a spatial segment $\ell$ and its complement experiences a phase transition as the length of the segment approaches a critical value $\ell_c$.  This behavior was interpreted as a confinement/deconfinement transition. 
 
One wonders, however, whether there is an alternative route that enables us to directly study the entanglement entropy, and other information-theoretic quantities, in confining gauge theories and examine their behavior near the deconfinement transition without the need to invoke the gauge/gravity duality or facing the ambiguities of the lattice formulation of gauge theory. In the present work we show that the answer to this question is affirmative. 

We study Yang-Mills theory compactified on a small spatial circle $\mathbb S_L^1$ and considered at temperatures near the deconfinement transition. The center of the theory is stabilized by means of deformations or by adding fermions obeying periodic boundary conditions along the circle. In the Euclidean setup  we say that the theory lives on $\mathbb R^2\times \mathbb T^2$, where  the two-torus $\mathbb T^2=\mathbb S_L^1\times \mathbb S_\beta^1$ and $\mathbb S_\beta^1$ is the thermal circle. This class of theories is adiabatically connected to Yang-Mills on $\mathbb R^4$ as we decompactify  $\mathbb S_L^1$, see, e.g., \cite{Unsal:2007jx,Poppitz:2012sw,Poppitz:2012nz,Anber:2014lba,Anber:2017tug}. For small enough $\mathbb S_L^1$ the theory is weakly coupled and dual to an XY-spin model with $\mathbb Z_p$ symmetry-preserving perturbations \cite{Anber:2011gn}.  The connection between the XY-spin model and Yang-Mills on $\mathbb R^2\times \mathbb T^2$ was made by mapping the partition functions of both theories to a  multi-component (dual) electric-magnetic Coulomb gas \cite{Kadanoff:1978ve,Kogan:2002au}\footnote{Also,  there have been attempts to describe the thermal dynamics of QCD on $\mathbb R^3\times \mathbb S_\beta^1$ as a plasma of classical electric and magnetic charges, see, e.g., \cite{Liao:2006ry}.}. The duality can also be derived more rigorously using the heat kernel methods in the presence of a non-trivial holonomy \cite{Anber:2013xfa}. Perturbations and vortices in the spin system map to magnetic charges (monopole- or bion-instantons) and electrically charged W-bosons in field theory (or vice versa, depending on the duality frame). Unlike the Svetitsky and Yaffe classification of the deconfinement transition \cite{Svetitsky:1982gs}, which is based on modeling the center symmetry of the gauge group using a scalar field theory, the gauge theory/XY-spin model duality is an exact mapping between both sides of the duality, at least within the validity of the Coulomb gas as an effective field description of gauge theory.

In fact, it can be shown that there exists two equivalent XY-spin descriptions of the gauge theory, which are the T-dual of each other. Moreover, an XY-spin model with $\mathbb Z_p$ symmetry-preserving perturbations is equivalent to a dual Sine-Gordon model, which again can be shown via the use of the dual Coulomb gas. This furnishes a web of dualities between Yang-Mills on a torus, XY-spin models, and dual Sine-Gordon models.

We exploit this web of dualities to study the entanglement entropy and mutual information in various flavors of Yang-Mills on $\mathbb R^2\times \mathbb T^2$. In particular, we consider Yang-Mills with center-preserving deformations in the absence and presence of fundamental fermions. We also consider a third example where we preserve the center by adding adjoint fermions obeying periodic boundary conditions along $\mathbb S^1_L$. In the three examples we study the entanglement entropy in the  dual Sine-Gordon model and show that this quantity exhibits a maximum at the transition/crossover temperature. 

Next, we study the R\'enyi mutual information (RMI) in the XY-spin models with $\mathbb Z_p$-preserving perturbations. We achieve this by considering a lattice version of the model and perform the computations using the replica trick and  Monte Carlo simulations. The advantage of the lattice XY-spin model over the lattice formulation of gauge theories is that the former does not suffer from ambiguities related to factorization of the Hilbert space.  We find that RMI follows the area law scaling, with subleading corrections, and its finite size scaling exhibits a clear crossing at the critical temperature, which is consistent with the location of the discontinuity of the magnetic susceptibility.  We observe this behavior in Yang-Mills with deformation and with adjoint fermions, while adding fundamental fermions washes out this effect. 

Our calculations are the first examples of using the entanglement entropy and RMI to probe phase transformations in weakly coupled confining gauge theories.   

This work is organized as follows. In Section \ref{Theory and formulation} we introduce our construction and review the main perturbative and nonperturbative ingredients of the theory. Since this class of theories have been extensively studied over the past decade, we keep our discussion brief. The interested reader can refer to a vast literature for more details.  In Section \ref{Finite temperature effects: the dual Coulomb gas, XY-spin model, dual Sine-Gordon model, and deconfinement} we consider the theory at temperatures near the transition point and construct the dual Coulomb gas. We also show the equivalence between this gas and XY-spin and dual Sine-Gordon models. The T-duality of the XY-spin model is also elucidated in this section. A reader who is familiar with the web of dualities we discuss in the present work can skip directly to Section \ref{Entanglement entropy and mutual information}, which is devoted to the study of the entanglement entropy and mutual information. After a brief introduction to these tools, we study the behavior of the former quantity in the dual Sine-Gordon model via analytical techniques. Next, we turn to a lattice version of the XY-spin model and use the replica trick and Monte Carlo simulations to numerically calculate RMI. Our numerical results are presented in Section \ref{Monte Carlo simulations}. We end with a discussion and future directions in Section \ref{Discussion and future directions}.

%%%%%%%%%%%%%%%%%%%%%%%%%%%%%%%%%%%%%
\section{Theory and formulation}
\label{Theory and formulation}
%%%%%%%%%%%%%%%%%%%%%%%%%%%%%%%%%%%%%

We consider $SU(2)$ Yang-Mills theory compactified over a circle $\mathbb S^1_L$ with circumference $L$, which is taken to be much smaller than the strong coupling scale, i.e., $L\Lambda_{QCD}\ll 1$:
\begin{eqnarray}
S_{SU(2)}=\int_{\mathbb R^3\times \mathbb S^1_L} \frac{1}{2g^2}\mbox{tr}_F\left[F_{MN}F_{MN} \right]\,,
\end{eqnarray}
where $g$ is the $4$-D coupling constant. We say that the theory lives on $\mathbb R^3\times \mathbb S^1_L$, and we take the circle in  the $x_3$ direction. In this work we use the upper case Latin letters to denote four dimensional quantities, $M,N=0,1,2,3$, while we use Greek alphabets to denote quantities on $\mathbb R^3$, i.e., $\mu,\nu=0,1,2$. We also adopt the normalization $\mbox{tr}_F\left[\tau^a\tau^b\right]=\delta^{ab}$, where $\{\tau^a\}$ are the $SU(2)$ color matrices. This amounts to using the fundamental weight $\omega=\frac{1}{\sqrt 2}$ and the root $\alpha=\sqrt{2}$.  Since the circle is small, the theory is in its weakly coupled regime and we can perform reliable perturbative/semi-classical analysis. However, in this regime the theory breaks its center symmetry. In order to restore the center, one needs to modify the theory in one of two ways.

The first option is to add a double-trace deformation to the $3$-dimensional reduced theory \cite{Unsal:2008ch}:
\begin{eqnarray}
\Delta S=\int_{\mathbb R^3} \frac{a}{L^3}|\Omega|^2\,,
\label{DTD}
\end{eqnarray}
where $a$ is a dimensionless coefficient that has to be taken large enough to win over the gauge field fluctuations that destabilize the center. The quantity $\Omega=\mbox{tr}_F\left[e^{i \oint_{\mathbb S^1_L} A_3}\right]=\mbox{tr}_F\left[e^{i LA_3^3\tau_3}\right]$ is the  fundamental Polyakov loop wrapping around the circle, and we have chosen the gauge field $A_3$ to lie along the third direction in the color space\footnote{Such a choice can always be made using an $SU(2)$ global transformation.}. This theory is known as deformed Yang-Mills, or dYM for short.  

 The other method we can use in order to preserve the center symmetry is to add fermions in the adjoint representation of the gauge group and give them periodic boundary conditions along $\mathbb S^1_L$. In this regard, this theory is distinct from thermal field theories, where the fermions obey anti-periodic boundary conditions. Upon dimensionally reducing the theory to $3$-D, we integrate out a tower of Kaluza-Klein excitations of fermions and gauge fields \cite{Gross:1980br}. This gives rise to the effective action \cite{Unsal:2007jx}
\begin{eqnarray}
\Delta S=\int_{\mathbb R^3}\frac{2(-1+n_f)}{\pi^2 L^3}\sum_{n=1}^\infty \frac{1}{n^4}|\Omega^n|^2\,,
\label{Full potential}
\end{eqnarray}   
where $n_f$ is the number of the massless Weyl fermions\footnote{For asymptotically free theory we take $n_f\leq 5.5$. The case $n_f=1$ corresponds to super Yang-Mills (SYM), and we refrain from discussing it in this work. For extensive works on SYM on $\mathbb R^3\times\mathbb S^1_L$ see \cite{Davies:2000nw,Anber:2017ezt,Anber:2014lba,Anber:2013doa}.}. We call this class of theories QCD(adj). In fact, one can also use massive adjoint fermions with masses $m\leq L^{-1}$ in order to stabilize the center. This is, however, effectively equivalent to adding a double-trace deformation. In this work we limit our treatment to dYM and QCD(adj) with massless fermions. The reader can refer to the following list of references \cite{Dunne:2018hog,Anber:2017ezt,Cherman:2017dwt,Anber:2017tug,Cherman:2017tey,Tanizaki:2017qhf,Poppitz:2017ivi,Aitken:2017ayq,Anber:2017rch,Cherman:2016jtu,Anber:2015wha}, which examined different aspects of QCD-like theories on a circle.  

In order to examine the effect of fundamental matter on the deconfinement transition, we also consider dYM in the presence of fundamental Dirac fermions\footnote{The maximum number of Dirac fermions one can add before losing the asymptotic freedom of the theory is $11$.} obeying periodic boundary conditions along $\mathbb S^1_L$. Indeed, the addition of fundamental fermions will push the theory towards breaking the center symmetry. However, we can always counter act this effect by taking the coefficient $a$ in (\ref{DTD}) to be large enough. We call this theory deformed Yang-Mills with fundamentals, or dYM(F) for short. 

%%%%%%%%%%%%%%%%%%%%%%%%%%%%%%%%%%%%%%%%
\subsection{The perturbative spectrum}
%%%%%%%%%%%%%%%%%%%%%%%%%%%%%%%%%%%%%%%%

In this section we analyze the perturbative spectrum of each of the three theories we considered above: dYM, dYM(F), and QCD(adj). Upon dimensionally reducing these theories to $3$-D, a nonzero vacuum expectation value (vev) of $A_3$ develops, and therefore, the gauge group $SU(2)$ breaks spontaneously to $U(1)$. One can use (\ref{DTD}) or (\ref{Full potential}) to show easily that the vev of $A_3$ is given by $LA_3^3=\frac{\pi}{\sqrt{2}}$. As we mentioned above, the vev respects the center symmetry because of either adding a deformation to the theory or using adjoint fermions. In $3$-D the photon has a single degree of freedom, and hence, we can go to a dual picture where we can describe it using a single scalar $\sigma$ via the duality relation $F^3_{\mu\nu}=\frac{g^2}{4\pi L}\partial_\alpha\sigma \epsilon_{\alpha\mu\nu}$. Then, the photon's kinetic energy reads 
\begin{eqnarray}
{\cal L}_{U(1)}=\frac{g^2}{16\pi^2 L}\left(\partial_\mu\sigma\right)^2\,.
\end{eqnarray}
$\sigma$ is a compact scalar valued in $\mathbb R/2\pi\omega$, or in other words, we impose the identification $\sigma\sim \sigma+\frac{2\pi}{\sqrt 2}$. The gauge field components that are perpendicular to the third color direction acquire a mass $M_W=\frac{\pi}{L}$ and become charged under the unbroken $U(1)$; namely these are the electrically charged W-bosons with electric charges valued in the root system\footnote{There are also higher Kaluza-Klein modes of W-bosons, which are much heavier than $M_W$, and hence, we neglect them in our treatment.}. In particular, the charges are $Q_W=\pm\sqrt{2}$. Upon adding fundamental fermions to dYM, i.e., for dYM(F), one finds that the fermions acquire a mass $M_F=A^3_3\omega=\frac{\pi}{2L}$ and charges $\pm\omega=\pm\frac{1}{\sqrt 2}$ under $U(1)$. The fundamental fermions are lighter than the W-bosons, and hence, we expect that they will dominate the dynamics in dYM(F). Finally, upon adding adjoint fermions we find that their component along the third direction is massless and uncharged under $U(1)$, and thus, it does not participate  in the dynamics of our theory. The other two components acquire a mass $M_{adj}=M_W=\frac{\pi}{L}$ and charges $Q_{adj}=Q_W=\pm\sqrt{2}$. In this regard, they are indistinguishable from  W-bosons on the classical level. We will see below that near the deconfinement transition all particles behave classically and one needs not distinguish between adjoint fermions and W-bosons. This completes the discussion of the perturbative spectrum. For more details the reader should consult \cite{Anber:2011de,Poppitz:2017ivi}.

%%%%%%%%%%%%%%%%%%%%%%%%%%%%%%%%%%%%%%%%%%%
\subsection{The nonperturbative spectrum}
%%%%%%%%%%%%%%%%%%%%%%%%%%%%%%%%%%%%%%%%%%%

In addition to the perturbative sector, our theories admit nonperturbative saddles. These are monopole- and bion-instantons. The monopole-instantons are a direct sequence of the nontrivial second homotopy group. In fact, in a center-symmetric vacuum we have two types of monopole-instantons with the exact same action $S_M=\frac{4\pi^2}{g^2}$ and charges $Q_M=\pm \sqrt{2}$: the normal BPS ('t Hooft-Polyakov) and twisted (first Kaluza-Klein) monopoles\footnote{There is an infinite tower of these monopoles. However, only the ones with the smallest action modify the IR dynamics of the theory.}, see \cite{Kraan:1998sn,Lee:1997vp}.  In dYM both types of monopoles participate in the dynamics; the proliferation of these monopoles causes the theory to develop a mass gap and the electric charges to confine. This is the celebrated Polyakov's confining mechanism \cite{Polyakov:1976fu,Anber:2013xfa}. The presence of fundamental fermions in dYM(F) modifies this picture slightly. While the twisted monopole does not get affected by the presence of fermions, the BPS monopole will acquire a single fermionic zero mode. This can be envisaged either by solving the Dirac's equation in the background of a single monopole \cite{Jackiw:1975fn,Csaki:2017cqm} or from the Callias' index \cite{Callias:1977kg,Nye:2000eg,Poppitz:2008hr}. Therefore, only one type of monopoles participates in the generation of the mass gap in dYM(F).  

The effect of monopoles can be taken into account in the partition function by inserting the vertex $e^{-\frac{4\pi^2}{g^2}e^{\pm i \sqrt{2}\sigma(x)}}$ at arbitrary spacetime points. Since $g\ll 1$, we find that the mean free path between the monopoles $\sim L e^{\frac{4\pi^2}{3g^2}}$ is much larger than their core radius ($\sim L$). This is the dilute gas limit, and thus, one can perform a reliable summation of  the monopole contribution to the partition function.  The resulting effective IR Lagrangian of both dYM and dYM(F) takes the form
\begin{eqnarray}
{\cal L}_{eff}=\frac{g^2}{16\pi^2 L}\left[\left(\partial_\mu\sigma\right)^2+m_\sigma^2\cos(\sqrt{2}\sigma)\right]\,,
\label{eff Lagrangian of dYM and dYM(F)}
\end{eqnarray}  
where $m_\sigma\sim \frac{e^{-\frac{4\pi^2}{g^2}}}{L}$ is the mass gap (monopole fugacity). From the discussion above we conclude that the fugacity of dYM is twice that of dYM(F).

The adjoint fermions in QCD(adj) makes the magnetic sector more complex. The index theorem indicates that both types of monopoles have two fermionic zero modes, and hence, they cannot participate directly in generating a mass gap. However, correlated monopole events made of a single BPS and a single twisted monopoles can form. The resulting molecules are dubbed magnetic-bions \cite{Unsal:2007jx,Anber:2011de}. They carry twice the action and twice the charge of a single monopole-instanton: $S_B=\frac{8\pi^2}{g^2}, Q_B=\pm 2\sqrt{2}$. The IR Lagrangian takes the form
\begin{eqnarray}
{\cal L}_{eff}=\frac{g^2}{16\pi^2 L}\left[\left(\partial_\mu\sigma\right)^2+m_\sigma^2\cos(2\sqrt{2}\sigma)\right]\,,
\label{Lagrangian for QCDadj}
\end{eqnarray}  
where $m_\sigma\sim \frac{e^{-\frac{8\pi^2}{g^2}}}{L}$ is the mass gap (bion fugacity) of QCD(adj). 

Finally, since both dYM and QCD(adj) have a $\mathbb Z_2^C$ center symmetry, an order parameter that transforms nontrivially under this symmetry can be used to distinguish between different phases. This is the Polyakov loop that wraps around the time circle. In addition, QCD(adj) enjoys a $\mathbb Z_2^{d\chi}$ discrete  chiral symmetry, which is broken in the low temperature regime \cite{Unsal:2007jx}. We discuss the thermal properties of our systems in the next section.

%%%%%%%%%%%%%%%%%%%%%%%%%%%%%%%%%%%%%%%%%%%%%%%%%%%%%%%%%%%%%%%%%%%%%%%%%%%%%%%%%%%%%%%%%%%%%%%%%%%%%%%%%%%%%%%%%%%%%
\section{Finite temperature effects: the dual Coulomb gas, XY-spin model, dual Sine-Gordon model, and deconfinement}
\label{Finite temperature effects: the dual Coulomb gas, XY-spin model, dual Sine-Gordon model, and deconfinement}
%%%%%%%%%%%%%%%%%%%%%%%%%%%%%%%%%%%%%%%%%%%%%%%%%%%%%%%%%%%%%%%%%%%%%%%%%%%%%%%%%%%%%%%%%%%%%%%%%%%%%%%%%%%%%%%%%%%%%%

%%%%%%%%%%%%%%%%%%%%%%%%%%%%%%%%%%%%%%%%
\subsection{The dual Coulomb gas}
%%%%%%%%%%%%%%%%%%%%%%%%%%%%%%%%%%%%%%%%

%%%%%%%%%%%%%%%%%%%%%%%%%%%%%%%%%%%%%%%%%%%%%%%%%%%%%%%%%%%%%%%%%%%%%%
\begin{figure}[t] %  figure placement: here, top, bottom, or page
   \centering
   \includegraphics[width=6in]{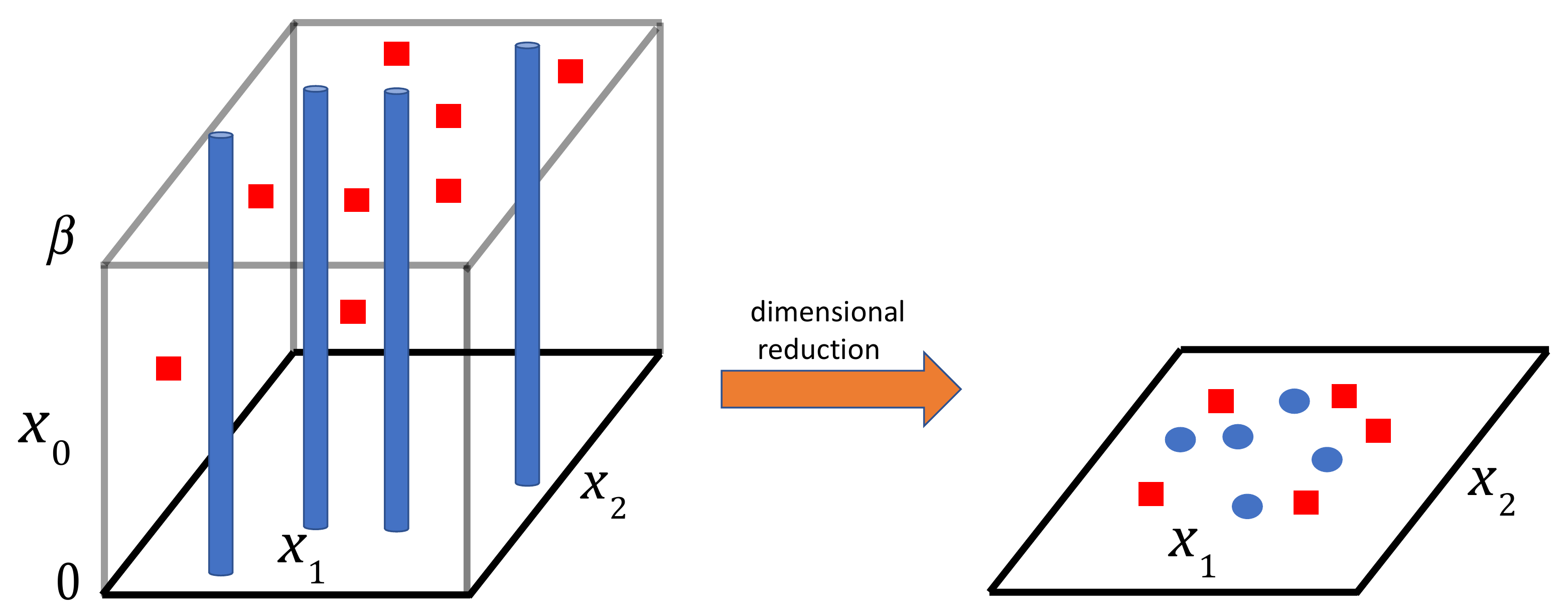} 
   \caption{The$3$-D system at finite temperature consists of W-bosons (represented by blue vortices) and instantons (represented by red square). The W-bosons are genuine particles, and hence, they trace worldlines (these are the blue vortices). In addition to the W-bosons we have instantons, which appear as localized objects in the Euclidean space. At finite temperature we identify the time-direction $x_0=0\sim \beta$ and the worldlines of W-bosons become closed circles.  Near the transition we can neglect all the Matsubara modes keeping only the zero mode; the system becomes effectively a 2-D Coulomb gas.}
	\label{Coulomb gas}
\end{figure}
%%%%%%%%%%%%%%%%%%%%%%%%%%%%%%%%%%%%%%%%%%%%%%%%%%%%%%%%%%%%%%%%%%%%%%%%%%%%

In this section we analyze the competing degrees of freedom as we consider our theory at a finite temperature. To this end we formulate dYM, dYM(F), and QCD(adj) on $\mathbb R^2\times \mathbb S^1_L \times \mathbb S^1_\beta$, where $\mathbb S^1_\beta$ is the time (thermal) circle. Thus, the fermions obey anti-periodic boundary conditions along $\mathbb S^1_\beta$, while they still obey periodic boundary conditions along $\mathbb S^1_L$. The temperature $T=\frac{1}{\beta}$ is assumed to be much smaller than the W-boson mass, i.e., $\beta\gg L$, and hence, we are far from the point of $SU(2)$ symmetry restoration (melting point of the W-bosons).   At this range of temperatures, both W-bosons and heavy fermions participate in the dynamics of the theory. Their fugacities will follow the Boltzmann's distribution:
\begin{eqnarray}
\xi_W\sim \frac{T}{L}e^{-\frac{M_W}{T}}\,,\quad \xi_F \sim \frac{T}{L}e^{-\frac{M_F}{T}}\,.
\end{eqnarray}
At temperatures close to the deconfinement transition, which will be shown to be much smaller than the melting temperature of W-bosons, only the  massless mode along $\mathbb S^1_\beta$ is important. Therefore, our theories can be dimensionally reduced to $2$-D and effectively we have a gas of magnetic (monopoles or bions) and electric (W-bosons or charged fermions) charges; see Figure \ref{Coulomb gas}. The fugacities of magnetic monopoles and bions are \cite{Anber:2011gn}:
\begin{eqnarray}
\xi_M\sim \frac{e^{-\frac{4\pi^2}{g^2}}}{L^3T}\,,\quad \xi_B \sim \frac{e^{-\frac{8\pi^2}{g^2}}}{L^3T}\,.
\end{eqnarray}
\begin{table}
\center
\begin{tabular}{|c|c|c|c|c|c|c|c|c|}
\hline
 & $Q_F$ & $\xi_F$ & $Q_W$ & $\xi_W$ & $Q_M$ & $\xi_M$ & $Q_B$ & $\xi_B$\\
\hline
dYM & \textemdash & \textemdash & $\pm\sqrt 2$ & $\frac{T}{L}e^{-\frac{M_W}{T}}$ & $\pm\sqrt 2$ & $\frac{e^{-\frac{4\pi^2}{g^2}}}{L^3T}$ & \textemdash&\textemdash\\
\hline
dYM(F) & $\pm\frac{1}{\sqrt 2}$ & $\frac{T}{L}e^{-\frac{M_F}{T}}$ & $\pm\sqrt 2$ & $\frac{T}{L}e^{-\frac{M_W}{T}}$ & $\pm\sqrt 2$ & $\frac{e^{-\frac{4\pi^2}{g^2}}}{L^3T}$ & \textemdash & \textemdash\\
\hline
QCD(adj) & \textemdash & \textemdash & $\pm\sqrt 2$ & $\frac{T}{L}e^{-\frac{M_W}{T}}$ & \textemdash & \textemdash & $\pm2\sqrt 2$ & $\frac{e^{-\frac{8\pi^2}{g^2}}}{L^3T}$\\\hline
\end{tabular}
\caption{Charges and fugacities of the electric and magnetic components in each theory.}
\label{Table of charges}
\end{table}
 This gas has been considered before  in \cite{Anber:2011gn,Teeple:2015wca} in great details. Here, we only summarize the final picture. First, any electrically charged objects in $2+1$-D will experience logarithmic potential, which is also true after compactifying the time direction \cite{Dunne:2000vp}:
\begin{eqnarray}
V(Q_{e_1},Q_{e_2})=-\frac{g^2  Q_{e_1}Q_{e_2}}{4\pi LT} \log T|\bm R_1-\bm R_2|\,.
\end{eqnarray} 
The potential between magnetically charged instantons in $3$-D Euclidean space follows the inverse square law. Upon dimensionally reducing the theory to $2$-D we obtain the logarithmic potential:
\begin{eqnarray}
V(Q_{m_1},Q_{m_2})=-\frac{4\pi LTQ_{m_1}Q_{m_2}}{g^2} \log T|\bm R_1-\bm R_2|\,.
\end{eqnarray}
In addition, magnetic and electric charges will experience Aharonov-Bohm interaction:
\begin{eqnarray}
V(Q_{e},Q_{m})=i 2Q_e Q_m\Theta(\bm R_e-\bm R_m)\,,
\end{eqnarray}
where $\Theta$ is the angle between the vector $\bm R_e-\bm R_m$ and the $x_2$-axis.

The mean free path between the various components of the gas is exponentially larger than their core radius ($\sim L$). For example, the mean free path between W-bosons or fermions is $l_{\mbox{mfp}}\sim Le^{\frac{M_{W,F}}{3T}}$. We show below that the transition temperature $T_c \sim \frac{g^2}{\pi L}$, and therefore, $l_{\mbox{mfp}}\sim Le^{\frac{\pi^2}{g^2}}\gg L$. Also, near the transition temperatures the momentum of W-bosons or fermions is $p\sim \sqrt{M_{W,F}T_c}$ and the corresponding De Broglie wavelength, $\lambda\sim \frac{L}{g^2}$, is much smaller than the mean free path. We conclude that our Coulomb gas is classical in nature.

At this stage, with the help of Table (\ref{Table of charges}), we are ready to write down the dual (multi-component) Coulomb gas Hamiltonian for our theories\footnote{It is dual in the sense that both electric and magnetic components are present in the gas.}.    In the following we will use subscripts with upper case Latin letters to denote  W-bosons, lower case Latin letters to denote monopoles or bions, and Greek letters to denote Fundamental fermions. The dual Coulomb gas of dYM contains W-bosons and magnetic monopoles. Its Hamiltonian reads:
\begin{eqnarray}
\nonumber
-\beta H_{dYM}&=&\frac{8 \pi LT}{g^2}\sum_{a>b}q_aq_b\log T|\bm R_a-\bm R_b|+\frac{g^2}{2\pi LT}\sum_{A>B}q_Aq_B\log T|\bm R_A-\bm R_B|\\
&&+i2\sum_{A,a}q_Aq_a \Theta\left(\bm R_a-\bm R_A\right)\,,
\label{Coulomb gas for dYM}
\end{eqnarray} 
where we use $\{q_a,q_A=\pm 1\}$ to denote the positive and negative charges. The dual Coulomb gas of dYM(F) contains the fundamental fermions as an extra component:
\begin{eqnarray}
\nonumber
-\beta {H}_{dYM(F)}&=&\frac{8\pi LT}{g^2}\sum_{a>b}q_aq_b\log T\left|\bm R_a-\bm R_b\right|+\frac{g^2}{2\pi LT}\sum_{A>B}q_Aq_B \log T\left|\bm R_A-\bm R_B\right|\\
\nonumber
&&+\frac{g^2}{8\pi LT}\sum_{\alpha>\beta}q_\alpha q_\beta \log T\left|\bm R_\alpha-\bm R_\beta\right|+\frac{g^2}{4\pi LT}\sum_{\alpha,A}q_\alpha q_A \log T\left|\bm R_\alpha-\bm R_A\right|\\
&&+i2\sum_{A,a}q_Aq_a\Theta(\bm R_a-\bm R_A)+i\sum_{\alpha,a} q_\alpha q_a\Theta(\bm R_a -\bm R_\alpha)\,.
\label{Coulomb gas for dYMF}
\end{eqnarray}
In fact, since the fundamental fugacity is exponentially larger than that of the W-bosons  (the fundamental fermions are much lighter than the W-bosons), we can neglect the latter in the Coulomb gas. Finally, the dual Coulomb gas of QCD(adj) is
\begin{eqnarray}
\nonumber
-\beta H_{QCD(adj)}&=&\frac{32 \pi LT}{g^2}\sum_{a>b}q_aq_b\log T|\bm R_a-\bm R_b|+\frac{g^2}{2\pi LT}\sum_{A>B}q_Aq_B\log T|\bm R_A-\bm R_B|\\
&&+i4\sum_{A,a}q_Aq_a \Theta\left(\bm R_a-\bm R_A\right)\,.
\label{Coulomb gas for QCDadj}
\end{eqnarray} 
Here, we note that both W-bosons and the heavy adjoint fermions are treated on equal footing since they are indistinguishable classically: they have the same fugacity and we use the same letter $A$ to denote both of them.    

The grand canonical partition function of the dual Coulomb gas is given by an arbitrary sum over all species weighted by their fugacities:
\begin{eqnarray}
\nonumber
{\cal Z}&=&\sum_{k=0}^\infty\int d^2R_{A_1}\int d^2R_{A_2}...\int d^2R_{A_k}\left(\xi_e\right)^k\\
&&\times \sum_{p=0}^\infty\int d^2R_{a_1}\int d^2R_{a_2}...\int d^2R_{a_p}\left(\xi_m\right)^p  e^{-\beta H}\,,
\label{partition function of the Coulomb gas}
\end{eqnarray}
where $\xi_e$ and $\xi_m$ are respectively the electric and magnetic fugacities. The competition between the different degrees of freedom of the gas determines the nature of phase transition or crossover as we dial the temperature. Also, different theories enjoy different discrete symmetries, as we discuss below. These symmetries get broken/restored in different phases.

%%%%%%%%%%%%%%%%%%%%%%%%%%%%%%%%%%%%%%%%%%%%%%%%%%%%%%%%%%%%%%%%%%%
\subsection{Equivalence between dual Coulomb gas and XY-spin model}
\label{Equivalence between dual Coulomb gas and XY-spin model}
%%%%%%%%%%%%%%%%%%%%%%%%%%%%%%%%%%%%%%%%%%%%%%%%%%%%%%%%%%%%%%%%%%%

The $2$-D dual Coulomb gas described by the partition function (\ref{partition function of the Coulomb gas}) and the Hamiltonians (\ref{Coulomb gas for dYM}), (\ref{Coulomb gas for dYMF}), or (\ref{Coulomb gas for QCDadj}) can be mapped to a $2$-D XY-spin model. Such equivalence was rigorously proven in various previous works, see e.g., \cite{Kadanoff:1978ve,Jose:1977gm}. Here we demonstrate this equivalence by showing that the partition function of the XY-spin model reproduces the grand canonical partition function of the dual Coulomb gas. 

The XY-spin model action is given by
\begin{eqnarray}
S[K,G_p,p]=\int d^2x\frac{K}{4\pi}\left(\partial_\mu \theta\right)^2-2G_p \cos\left(p\theta\right)\,,
\label{general S XY spin model}
\end{eqnarray}
where $\theta$ is a compact scalar field, i.e., $\theta\sim \theta+2\pi$, and $G_p \cos\left(p\theta\right)$, where $p \in \mathbb Z^+$,  are $\mathbb Z_p$ symmetry-preserving perturbations. The kinetic term is invariant under a $U(1)$ symmetry, $\theta\rightarrow \theta+c$, which is explicitly broken by the perturbations down to a $\mathbb Z_p$ subgroup: $\theta\rightarrow \theta +\frac{2\pi}{p}$. The partition function reads:
\begin{eqnarray}
Z[K,G_p,p;H_w,w]=\int {\cal D}\theta e^{-S[K,G_p,p]}\,.
\label{general Z XY spin model}
\end{eqnarray}
The meaning of $K,G_p,p$ as arguments of $Z$ is evident, while the meaning of $H_w$ and $w$ is not yet clear. In the following we clarify this meaning and elucidate the connection between the XY-spin model and dual Coulomb gas. 

To this end we write $2G_p\cos(p\theta)$ in (\ref{general S XY spin model}) as $G_p\left(e^{ip\theta}+e^{-ip\theta}\right)$ and expand the action as a series in $G_p$:
\begin{eqnarray}
\nonumber
e^{\int d^2 x 2G_p\cos(2\theta_p)}&=&\sum_{k\geq 0} \frac{(2G_p)^k}{k!}\left(\int d^2 x \frac{e^{ip\theta(\bm R)}+e^{-ip\theta(\bm R)}}{2}\right)^k\\
&=&\sum_{n\geq0} \sum_{q_J=\pm 1}\frac{(G_p)^{2n}}{\left(n!\right)^2}\prod_{J=0}^{2n}\int d^2 x_J e^{iq_J\theta(\bm R_J)}\,.
\label{intermediate step in xy to Coulomb gas}
\end{eqnarray}
$q_J$ is interpreted as the charge of a particle inserted at location $\bm R_J$. In other words, the insertion of the operator $e^{iq_J\theta(\bm R_J)}$ creates a charge $q_A$ at position $\bm R_J$. This is the first step needed in order to recognize that the partition function of the XY-spin model can be rewritten as the grand canonical partition function of a collection of charged particles.  Notice that we have assumed  an equal number of positive and negative charges in going from the first to second line above. The neutrality of the total charge of the system, i.e., $\sum_{A} q_J=0$, is important in order to have a well defined partition function in $2$-D \cite{ZinnJustin:2002ru}. Next, we insert (\ref{intermediate step in xy to Coulomb gas}) into (\ref{general Z XY spin model}) to find that the Gaussian action $\int d^2 x (\partial_\mu\theta)^2$ is sourced by the charges located at $\bm R_J$. The resulting equation of motion of $\theta$ reads $\nabla^2\theta=-i\sum_{J} pq_J\delta^{(2)}(\bm R-\bm R_J)$. Since $\theta$ is a compact scalar, its most general solution contains vortices with arbitrary integer winding numbers $w=q_j$ located at arbitrary positions $\bm R_j$:
\begin{eqnarray}
\theta(\bm R)=-\frac{ip}{K}\sum_{J}q_J\log T |\bm R-\bm R_J|+\sum_{j} q_j\Theta(\bm R-\bm R_j)+\theta_0(\bm R)\,,
\label{general solution of theta}
\end{eqnarray}
where the temperature $T$ is an IR regulator that is introduced to make the argument of the log dimensionless \footnote{One can also introduce a UV cutoff for the same reason.}  and also for an obvious convenience, $\theta_0(\bm R)$ are periodic spin-wave fluctuations, and the vortices satisfy the neutrality condition $\sum_{a}q_j=0$. The creation of a vortex costs a certain amount of core energy which increases with the winding number. Therefore, the partition function (\ref{general Z XY spin model}) depends implicitly on the vortex winding number $w$ and its fugacity $H_w$. Finally, we substitute the solution (\ref{general solution of theta}) into  (\ref{general Z XY spin model}) and sum over an arbitrary number of vortices of charge $q=\pm w$ and fugacity $H_w$, to obtain 
\begin{eqnarray}
\nonumber
Z[K,G_p,p;H_w,w]&=&Z_0\sum_{m,q_j=\pm w}\sum_{n,q_J=\pm 1}\frac{G_p^{2n}}{(n!)^2}\frac{H_w^{2m}}{(m!)^2}\prod_{J=0}^{2n}\int d^2 x_J \prod_{j=0}^{2m}\int d^2 x_j\\
\nonumber
&\times& \exp\left[\sum_{J_1>J_2}\frac{p^2}{K}q_{J_1}q_{J_2} \log T|\bm R_{J_1}-\bm R_{J_2}|+\sum_{j_1>j_2}K q_{j_1}q_{j_2} \log T|\bm R_{j_1}-\bm R_{j_2}|\right.\\
&&+\left.ip\sum_{J,j}q_Jq_j\Theta(\bm R_j-\bm R_J)\right]\,,
\label{the xy coulomb gas equivalence}
\end{eqnarray}
where $Z_0$ is the partition function of the spin-wave fluctuations. The fugacity $H_w$ is an implicit parameter of the partition function (\ref{the xy coulomb gas equivalence}) since its precise value can't be determined apriori. In a UV regularization of the theory the value of $H_w$ is of the same order of magnitude of the cutoff scale, i.e., $H_w\sim \Lambda^2$ . For example, one can regularize the theory by putting it on a lattice to find $H_w \sim a^{-2}$, where $a$ is the lattice spacing, see \cite{Kadanoff:1978ve,Jose:1977gm,Wen:2004ym} for more details. This is exactly what we do in Section \ref{Mutual information from the XY-spin model on the lattice T duality}.

 It is important to emphasize that the subscripts $j$ and $J$ in (\ref{the xy coulomb gas equivalence}) can denote either the electrically or magnetically charged particles, with no preference at this point. The partition function (\ref{the xy coulomb gas equivalence}) is invariant under a $2\pi$ monodromy of $\Theta(\bm R_j-\bm R_J)$, and hence, the product $pq_jq_J\in \mathbb Z$. This completes the proof of the equivalence between the partition function of the XY-spin model and dual Coulomb gas.
 
The fact that $q_j$ and $q_J$ could denote either the electric or magnetic charges give us the freedom to write two equivalent XY-spin models for each theory we have at hand. In one model the electric charges are explicit while the magnetic charges are implicit, and vice versa for the second model. In fact, these two equivalent models are mapped to each other via a T-duality. In the following we elucidate this construction for dYM, dYM(F), and QCD(adj).

%%%%%%%%%%%%%%%%%%%%%%%%%%%%%%%%%%%%%%%%%%
\subsubsection*{{\bf dYM} and {\bf dYM(F)}}
%%%%%%%%%%%%%%%%%%%%%%%%%%%%%%%%%%%%%%%%%%
\begin{enumerate}
\item The partition function of a description where the 't Hooft-Polyakov magnetic monopoles are explicit is given by:
\begin{eqnarray}
\nonumber
&&S[K=\frac{g^2}{8\pi LT},G_1=\xi_M]=\int d^2x\frac{g^2}{32\pi^2 LT}\left(\partial_\mu \theta\right)^2-2\xi_m \cos\theta\,,\\
&&Z[K=\frac{g^2}{8\pi LT},G_1=\xi_M,p=1;H_1,H_2,w=\{1,2\}]=\int {\cal D}\theta e^{-S}\,.
\label{XY for dYM and dYMF}
\end{eqnarray}
This action can also be obtained from the $3-$D action (\ref{eff Lagrangian of dYM and dYM(F)}) after dimensionally reducing the theory to $2$-D and making the substitution $\sqrt 2 \sigma=\theta$. The operator $e^{\pm i\theta}$ creates an 't Hooft-Polyakov magnetic monopole with a unit charge, which is the lowest magnetic charge allowed in this description. The winding numbers $w=1,2$ are the fundamental and adjoints charges, receptively. Therefore, the vortices are the Dirac fermions ($w=1$) and W-bosons ($w=2$).  This can be easily envisaged from comparing the general Coulomb gas in (\ref{the xy coulomb gas equivalence}) with (\ref{Coulomb gas for dYM}) and (\ref{Coulomb gas for dYMF}). The fugacity of a unit winding vortex $H_1$ is naturally bigger than that of a vortex with twice the winding $H_2$ (or in other words, the core energy of $w=2$ is bigger than that of $w=1$). This exactly matches our expectation that the fugacity of the fundamental fermions is bigger than that of the W-Bosons. We conclude that $H_1=\xi_F, H_2=\xi_W$. Therefore, (\ref{XY for dYM and dYMF}) is a natural description of dYM(F). In order to remove the fermions from the description, and hence describe dYM, one has to exclude the unit-winding vortices. 

 The action in (\ref{XY for dYM and dYMF}) does not have an order parameter in the presence of $w=1$ vortices, and hence, one does not expect to see a phase transition in this system. In fact, it can be shown that this system is always in a gapped phase. 

\item In the dual description the W-bosons and fundamental fermions are explicit. The action and partition function take the form
\begin{eqnarray}
\nonumber
&&S[K=\frac{8\pi LT}{g^2},G_1=\xi_F,G_2=\xi_W]=\int d^2x\frac{2 LT}{g^2}\left(\partial_\mu \theta\right)^2-2\xi_F \cos\theta-2\xi_W\cos\left(2\theta\right)\,,\\
&&Z[K=\frac{8\pi LT}{g^2},G_1=\xi_F,G_2=\xi_W,p=\{1,2\};H_1,w=1]=\int {\cal D}\theta e^{-S}\,.
\label{dual XY for dYM and dYMF}
\end{eqnarray}
The operators $e^{\pm i\theta}$ and $e^{\pm 2i\theta}$ create a Dirac fermion and W-boson, respectively. The vortex with the lowest winding number $w=1$ corresponds to monopoles, i.e., $H_1=\xi_M$, as can be checked directly by comparing (\ref{the xy coulomb gas equivalence}) with (\ref{Coulomb gas for dYM}) and (\ref{Coulomb gas for dYMF}). Therefore, the action (\ref{dual XY for dYM and dYMF}) describes dYM(F) and in the special case $\xi_F=0$ it describes dYM. 

Setting $\xi_F=0$, i.e., for dYM, we find that the system enjoys a $\mathbb Z_2$ symmetry: $\theta \rightarrow \theta +\pi$. This is the $\mathbb Z_2^C$ zero-form center symmetry, which emerges upon compactifying the theory over $\mathbb S^1_\beta$.

\end{enumerate}

%%%%%%%%%%%%%%%%%%%%%%%%%%%%%%%%%%%%%%%%
\subsubsection*{{\bf QCD(adj)}}
%%%%%%%%%%%%%%%%%%%%%%%%%%%%%%%%%%%%%%%%
\begin{enumerate}
\item We start with the XY-model that explicitly accounts for the magnetic bions \cite{Anber:2011gn}:
\begin{eqnarray}
\nonumber
&&S[K=\frac{g^2}{8\pi LT},G_2=\xi_B]=\int d^2x\frac{g^2}{32\pi^2 LT}\left(\partial_\mu \theta\right)^2-2\xi_B \cos\left(2\theta\right)\,,\\
&&Z[K=\frac{g^2}{8\pi LT},G_2=\xi_B,p=2;H_2,w=2]=\int {\cal D}\theta e^{-S}\,.
\label{XY for QCDadj}
\end{eqnarray}
This is the direct generalization of (\ref{XY for dYM and dYMF}) from $p=1$ to $p=2$. The action (\ref{XY for QCDadj}) can be obtained from the $3$-D theory (\ref{Lagrangian for QCDadj}) after dimensionally reducing it to $2$-D and making the substitution $\sqrt 2 \sigma=\theta$. The operator $e^{\pm i 2 \theta}$ creates a magnetic bion, while the monopoles are not dynamical in this system. Instead, one can use the operator $e^{\pm i \theta}$ as an external probe.  The system allows for both $w=1,2$ vortices.  One needs, however, to suppress the $w=1$ vortices since they correspond to fundamental electric charges, while the allowed $w=2$ vortices are the adjoint fermions and W-bosons..  

\item In the dual description the action and partition function take the form
\begin{eqnarray}
\nonumber
&&S[K=\frac{8\pi LT}{g^2},G_4=\xi_W]=\int d^2x\frac{2 LT}{g^2}\left(\partial_\mu \theta\right)^2-2\xi_W\cos\left(4\theta\right)\,,\\
&&Z[K=\frac{8\pi LT}{g^2},G_4=\xi_W,p=4;H_1,w=1]=\int {\cal D}\theta e^{-S}\,.
\label{dual XY for QCDadj}
\end{eqnarray}
The operator $e^{\pm i4\theta}$ creates W-bosons, while $w=1$ vortices are the magnetic bions. An insertion of the operator $e^{\pm i 2\theta}$ creates a nondynamical fundamental electric charge, while the operator $e^{\pm i \theta}$ represents one-quarter the charge of W-bosons (such charge does not exist in $SU(2)$).
This action is invariant under a $\mathbb Z_4$ discrete symmetry: $\theta \rightarrow \theta+\frac{\pi}{2}$. QCD(adj) enjoys two types of discrete symmetries: the $\mathbb Z_2^{C}$ center and $\mathbb Z_2^{d\chi}$ discrete chiral symmetries. In fact, the action (\ref{dual XY for QCDadj}) enjoys the enhancement  $\mathbb Z_2^C\times \mathbb Z_2^{d\chi} \rightarrow \mathbb Z_4$.   
\end{enumerate}

%%%%%%%%%%%%%%%%%%%%%%%%%%%%%%%%%%%%%%%%%%%%%%%%%%%%%%%%
\subsection{The dual Sine-Gordon model and deconfinement}
\label{The dual Sine-Gordon model and deconfinement}
%%%%%%%%%%%%%%%%%%%%%%%%%%%%%%%%%%%%%%%%%%%%%%%%%%%%%%%%

Both the dual Coulomb gas and XY-spin model can also be mapped to the dual Sine-Gordon model \cite{Kovchegov:2002vi}. The latter can be used to estimate the critical temperature and universality class of the transition.  The dual Sine-Gordon action reads
\begin{eqnarray}
S=\int d^2x\frac{1}{2}\left(\partial_x\Phi \right)^2+\frac{1}{2}\left(\partial_x\chi \right)^2-i \partial_x\Phi\partial_\tau \chi-\frac{\alpha}{\kappa^2}\cos\left(\kappa\Phi\right)-\frac{\beta}{\rho^2}\cos\left(\rho\chi\right)\,,
\label{dual Sine-gordon model}
\end{eqnarray}
where both $\Phi$ and $\chi$ are noncompact scalars. The model enjoys a duality under the exchange $\Phi \leftrightarrow \chi$, $\kappa  \leftrightarrow \rho$, and $\alpha  \leftrightarrow \beta$.
The equivalence between (\ref{dual Sine-gordon model}) and the dual Coulomb gas can be shown by first rewriting the cosine terms in the form (\ref{intermediate step in xy to Coulomb gas}). The partition function of (\ref{dual Sine-gordon model})  then becomes
\begin{eqnarray}
\nonumber
{\cal Z}&=&\sum_{m,q_j=\pm w}\sum_{n,q_J=\pm 1}\frac{\left(\frac{-\alpha}{2\kappa^2}\right)^{2n}}{(n!)^2}\frac{\left(\frac{-\beta}{2\rho^2}\right)^{2m}}{(m!)^2}\prod_{J=0}^{2n}\int d^2 x_J \prod_{j=0}^{2m}\int d^2 x_j \left\langle \prod_{a=1}^k e^{i\kappa\Phi(\bm R_{J_a})} \prod_{b=1}^p e^{i\rho\chi(\bm R_{j_b})} \right\rangle_0\,,\\
\end{eqnarray}
where the average $\langle \,\, \rangle_0$ is taken with respect to $S_0$, which is the  massless free part of (\ref{dual Sine-gordon model}), and we also assumed the neutrality of the system. Using the expression of the free propagators (see \cite{Kovchegov:2002vi}): $\langle T \chi(\bm R)\chi(\bm 0) \rangle_0=\langle T \Phi(\bm R)\Phi(\bm 0) \rangle_0=-\frac{1}{2\pi}\log T|\bm R|$, $\langle T \chi(\bm R)\Phi(\bm 0) \rangle_0=\frac{i}{2\pi}\Theta(\bm R)$, and repeating the steps that lead from (\ref{intermediate step in xy to Coulomb gas}) to (\ref{the xy coulomb gas equivalence}),  we readily find
\begin{eqnarray}
\nonumber
\left\langle \prod_{a=1}^k e^{i\kappa\Phi(\bm R_{J_a})} \prod_{b=1}^p e^{i\rho\chi(\bm R_{j_b})} \right\rangle_0&=&\exp\left[\sum_{J_1>J_2}\frac{\kappa^2}{2\pi}\log T\left|\bm R_{J_1}-\bm R_{J_2}\right|+\sum_{j_1>j_2}\frac{\rho^2}{2\pi}\log T\left|\bm R_{j_1}-\bm R_{j_2}\right|\right.\\
&&\left. -i\sum_{J,j} \frac{\kappa\rho}{2\pi}\Theta\left(\bm R_{J}-\bm R_{j}\right)\right]\,.
\label{Sine Gordon Coulomb gas}
\end{eqnarray}

The scaling dimensions of $\cos\left(\kappa\Phi\right)$ and $\cos\left(\rho \Theta\right)$  can be obtained via the renormalization group equations  to find \cite{Boyanovsky:1988ge,Boyanovsky:1990iw}:
\begin{eqnarray}
\alpha(\mu)=\alpha_0\left(\frac{\mu}{\mu_0}\right)^{\Delta_\alpha-2}\,,\quad \beta(\mu)=\beta_0\left(\frac{\mu}{\mu_0}\right)^{\Delta_\beta-2}\,,
\label{RG equations}
\end{eqnarray} 
with $\Delta_\alpha\equiv \frac{\kappa^2}{4\pi}$ and $\Delta_\beta\equiv \frac{\rho^2}{4\pi}$ are the conformal dimensions of the corresponding cosine terms, $\mu_0$ is a UV energy scale, and $\alpha_0,\beta_0$ are the values of $\alpha,\beta$ at $\mu_0$. Therefore, the cosine terms are IR relevant for $\Delta_\alpha,\Delta_\beta<2$.

In the following we analyze each of our theories in the light of (\ref{dual Sine-gordon model}) and (\ref{RG equations}).

%%%%%%%%%%%%%%%%%%%%
\subsubsection*{dYM}
%%%%%%%%%%%%%%%%%%%

Comparing the Coulomb gases (\ref{Sine Gordon Coulomb gas}) and (\ref{Coulomb gas for dYM}) we find that $\Phi$ and $\chi$ are mapped to W-bosons and magnetic monopoles, respectively. Therefore, we have
\begin{eqnarray}
\kappa=\frac{g}{\sqrt{LT}}\,,\quad \rho=\frac{4\pi \sqrt{LT}}{g}\,,
\end{eqnarray}
and $\alpha$, $\beta$, are respectively the electric and magnetic fugacities. 
One can distinguish between three temperature ranges:
\begin{enumerate}
\item $T<\frac{g^2}{8\pi L}$. In this temperature range, and according to (\ref{RG equations}),  $\cos(\kappa \Phi)$ and  $\cos(\rho \chi)$ are IR irrelevant and relevant, respectively. Therefore, we expect the W-bosons to be confined in neutral pairs, while the vacuum is populated by a magnetic plasma. This is a magnetic discorded (gapped) phase. Therefore, one can integrate out the $\Phi$ field, which yields a $2$-D Sine-Gordon model of the magnetic plasma:
\begin{eqnarray}
{\cal L}_m=\frac{1}{2}\left(\partial_\mu\chi\right)^2-\frac{\beta}{\rho^2}\cos(\rho\chi)\,.
\label{magnetic sin Gordon}
\end{eqnarray}
\item $T>\frac{g^2}{2\pi L}$. This is the dual phase: the magnetic monopoles are confined in neutral pairs, while the W-bosons populate the vacuum. The system is in an electrically disordered (gapped) phase.  We integrate out the monopoles to obtain the 2-D Sine-Gordon model of the electric plasma:
\begin{eqnarray}
{\cal L}_e=\frac{1}{2}\left(\partial_\mu\Phi\right)^2-\frac{\alpha}{\kappa^2}\cos(\kappa\Phi)\,,
\label{electric sin Gordon}
\end{eqnarray}   
The Lagrangians (\ref{magnetic sin Gordon}) and (\ref{electric sin Gordon}) are the dual of each other. Thus, the dual Coulomb gas of dYM enjoys electric-magnetic duality. 
\item $\frac{g^2}{8\pi L}<T<\frac{g^2}{2\pi L}$. In this range both W-bosons and monopoles are relevant. A phase transition may occur in this range of temperatures. This can be envisaged by mapping the dual Sine-Gordon model to an effective fermionic theory via bosonization techniques \cite{Zuber:1976aa,Kogan:2002au}. An analysis of the fermionic theory \cite{Dunne:2000vp, Kovchegov:2002vi,Lecheminant:2002va} indicates that the system exhibits a $\mathbb Z_2$ Ising criticality at $T_c=\frac{g^2}{4\pi L}$. This is the self-dual point of the electric-magnetic duality.  
\end{enumerate}

%%%%%%%%%%%%%%%%%%%%%%%%%%%%
\subsubsection*{dYM(F)}
%%%%%%%%%%%%%%%%%%%%%%%%%%%%
The fugacity of the fundamental quarks is exponentially larger than that of W-bosons, see Table (\ref{Table of charges}). Therefore, W-bosons do not play an important rule in the IR dynamics and we ignore them in our treatment of dYM(F). Comparing (\ref{Coulomb gas for dYMF}) and  (\ref{Sine Gordon Coulomb gas}) we find 
\begin{eqnarray}
\kappa=\frac{g}{2\sqrt{LT}}\,, \quad \rho=\frac{4\pi}{g}\sqrt{LT}\,, 
\end{eqnarray} 
and $\alpha$, $\beta$ are the fugacities of fermions and monopoles, respectively. 
One can also divide the temperature into three ranges as in the case of dYM. The system is dominated by electric charges (fermions) at high temperatures, $T>\frac{g^2}{16\pi L}$, by monopoles at low temperatures $T<\frac{g^2}{4\pi L}$, and by both electric and magnetic charges in the range $\frac{g^2}{16\pi L}<T<\frac{g^2}{4\pi L}$. The system, however, is always in a gapped phase, and hence, it does not experience a phase transition. This can be shown explicitly by mapping the dual Sine-Gordon model with $\kappa=\frac{g}{2\sqrt{LT}}\,,\rho=\frac{4\pi}{g}\sqrt{LT}$ into a dimerized spin-$1/2$ antiferromagnetic Heisenberg chain in a staggered magnetic field \cite{Lecheminant:2002va}.  The system exhibits a crossover as it transforms from the electric to magnetic phases.

%%%%%%%%%%%%%%%%%%%%%%%%%%%%
\subsubsection*{QCD(adj)}
%%%%%%%%%%%%%%%%%%%%%%%%%%%%

The dual Coulomb gas of QCD(adj) gives
\begin{eqnarray}
\kappa=\frac{g}{\sqrt{LT}}\,,\quad \rho=\frac{8\pi}{g}\sqrt{LT}\,,
\end{eqnarray}
where $\alpha$ and $\beta$ are respectively mapped to the W-boson and magnetic bion fugacities. The theory again exhibits different behaviors in three different ranges of temperatures:
\begin{enumerate}
\item $T<\frac{g^2}{8\pi L}$. At low temperature the magnetic bions dominate the plasma and one integrates out the W-bosons to find that the system is described by the effective Lagrangian (\ref{magnetic sin Gordon}).
\item $T>\frac{g^2}{8\pi L}$. At high temperature the magnetic bions are confined and the W-bosons populate the system. Integrating out the magnetic charges, one finds that the system is described by the Lagrangian (\ref{electric sin Gordon}).
\item $T_c=\frac{g^2}{8\pi L}$. The theory is Gaussian (free) and exhibits a critical behavior exactly at this point, see \cite{Anber:2011gn, Lecheminant:2002va}.  This can be shown rigorously by mapping the dual Sine-Gordon model of QCD(adj) into an anisotropic version of the $su(2)_1$ Wess-Zumino-Novikov-Witten model with a current-current interaction \cite{Lecheminant:2002va}.
\end{enumerate}

%%%%%%%%%%%%%%%%%%%%%%%%%%%%%%%%%%%%%%%%%%%%%%%%%%%%%%%%
\section{Entanglement entropy and mutual information}
\label{Entanglement entropy and mutual information}
%%%%%%%%%%%%%%%%%%%%%%%%%%%%%%%%%%%%%%%%%%%%%%%%%%%%%%%%

In this work we are interested in using information-theoretic techniques to study gauge theories near the deconfinement transition. This works not only as an alternative point of view to Landau-Ginzburg criteria, but also as a new probe that may shed light on new properties of gauge theories. In this section we review essential concepts in information theory that are vital to our work.

%%%%%%%%%%%%%%%%%%%%%%%%%%%%%%%%%%%%%%%%%%%%%%%
\subsection{Elements of information theory}
%%%%%%%%%%%%%%%%%%%%%%%%%%%%%%%%%%%%%%%%%%%%%%%

Let a manifold ${\cal M}$ be bipartitioned into ${\cal A}$ and ${\cal B}$ such that ${\cal A} \cup {\cal B}={\cal M}$. Now, $\{x_i\} \in X$ and $\{y_i\}\in Y$  are two sets of random variables (a statistical field) with support on ${\cal A}$ and ${\cal B}$, respectively. For example, they can be two sets of disjoint spins on a lattice.   The expectation value of the random variable is given by $E(X)=\sum_{x \in \Phi}p(x)x $, and similarly for $E(Y)$. The function $p(x)$ is the probability distribution of the field, which could be, for example, the Boltzmann distribution.  The connected Green's function (or correlation function) is defined as $C(X,Y)=\sum_{x \in X, y \in Y}p(x, y)xy-E(X)E(Y)$, where $p(x,y)$ is the joint probability distribution between $X$ and $Y$. In the special case when $p(x,y)$ factors into $p(x)p(y)$, the correlation function vanishes. Whence, $C(X,Y)$ carries information about the correlation between different parts of the system. The disadvantage of $C(X,Y)$ is that it depends not only on the joint probability, but also depends explicitly  on the fields $X$ and $Y$, and therefore, it may overlook important mutual information between ${\cal A}$ and ${\cal B}$. This can happen, for example, if the values of $\{x_i\}$ and $\{y_i\}$ are small eventhough the two subspaces are highly correlated. While the fields themselves are not physical (one can always perform arbitrary transformations on the fields), the mutual information between ${\cal A}$ and ${\cal B}$, which is encoded in the joint probability between them, is physical. 

Fortunately enough, there is a quantity in the context of information theory that quantifies the correlation between two systems without making an explicit reference to the set of random variables (or fields). This quantity is the {\em mutual information}, which is defined via:
\begin{eqnarray}
I(X;Y)\equiv \sum_{x \in X, y\in Y}p(x,y)\log \left(\frac{p(x,y)}{p(x)p(y)}\right)\,.
\end{eqnarray}
It is easy to see that $I(X;Y)\geq 0$ and vanishes iff the joint probability factorizes: $p(x,y)=p(x)p(y)$. The mutual information measures the amount of information shared between ${\cal A}$ and ${\cal B }$. In other words, it quantifies how much information about ${\cal A}$ reduces the uncertainty about ${\cal B}$.

 The uncertainty of a physical quantity is quantified by entropy. In information theory this uncertainty is given by {\em Shannon's entropy}:
 \begin{eqnarray}
 S=-\sum_{i}p_i \log p_i\,.
 \end{eqnarray}
 Therefore, Shannon's entropy of ${\cal A}\cup{\cal B}$ reads
\begin{eqnarray}
S({\cal A}\cup{\cal B})\equiv -\sum_{x\in X, y\in Y} p(x,y)\log p(x,y)\,.
\end{eqnarray}  
The reduced entropy, $S({\cal A})$, is obtained by tracing out the degrees of freedom of ${\cal B}$:
\begin{eqnarray}
S({\cal A})= -\sum_{x \in X} p(x)\log p(x)\,,
\end{eqnarray}  
where $p(x)=\sum_{y \in Y}p(x,y)$ and a similar expression for $S({\cal B})$.  Then, one can show that \cite{2015arXiv150802595Z}:
\begin{eqnarray}
I(X;Y)=S({\cal A})+S({\cal B})-S({\cal A}\cup {\cal B}),
\end{eqnarray}
and in the case of perfect correlation (e.g., at zero temperature) both $I(X;Y)$ and $S(X)=S(Y)$ coincide. It can also be shown that $I(X;Y)$ is a non-increasing function as we eliminate parts of the system, i.e., under the renormalization group flow, see \cite{Nielsen:2011:QCQ:1972505}.  In a quantum system one replaces the probability $p$ with the density matrix $\rho$ and Shannon's entropy becomes $S({\cal A}\cup {\cal B})=-\mbox{tr}_{{\cal A}\cup {\cal B}}\left[\rho \log \rho\right]$, which is the von-Neumann entropy. The reduced entropy $S({\cal A})$ can be found in two step: first, one traces over system ${\cal B}$ to find the reduced density matrix  $\rho({\cal A})=\mbox{tr}_{{\cal B}}\rho$, and second, the reduced entropy is obtained via $S({\cal A})=-\mbox{tr}_{{\cal A}}\rho({\cal A})\log \rho({\cal A})$ . 

In many situations the direct calculations of Shannon's or von-Neumann entropies are plagued by many difficulties. In this case one instead can use the {\em generalized R\'enyi entropy}, which is defined as:
\begin{eqnarray}
S_n({\cal A}\cup {\cal B})=\frac{1}{1-n}\log\left(\sum_{x \in X, y\in Y}p^n(x,y)\right)\,,
\label{generalized Reny entropy}
\end{eqnarray}
and
\begin{eqnarray}
S_n({\cal A})=\frac{1}{1-n}\log\left(\sum_{x \in X}p^n(x)\right)\,,
\label{reduced generalized  Reny entropy}
\end{eqnarray}
such that Shannon's entropy is reproduced in the limit $S=\lim_{n \rightarrow 1} S_n$. Similarly, {R\'enyi mutual information} is given by the expression
\begin{eqnarray}
I_{n}(X;Y)=S_n({\cal A})+S_n({\cal B})-S_n({\cal A}\cup {\cal B})\,.
\label{Renyi MI}
\end{eqnarray}

Shannon's or von-Neumann entropies ($S({\cal A}\cup {\cal B})$, $S({\cal A})$, or $S({\cal B})$), or their R\'enyi generalization, are examples of entanglement entropies. Unlike thermodynamic entropy, which scales with the system size, entanglement entropy scales with area\footnote{The area law term is the leading term at zero temperature. At finite temperatures, though, there will also be a term that scales with volume.}. This behavior of entropy was first observed in the scaling of the black hole entropy with the area of the event horizon \cite{Bekenstein:1973ur, Bardeen:1973gs, tHooft:1984kcu}, and the concept was further developed by Takayanagi and Ryu in the AdS/CFT context \cite{Ryu:2006bv}. There has also been a plethora of applications of this concept in many-body physics and critical phenomena, see, e.g., \cite{PhysRevLett.90.227902}. 

\begin{figure}[t] %  figure placement: here, top, bottom, or page
   \centering
   \includegraphics[width=4in]{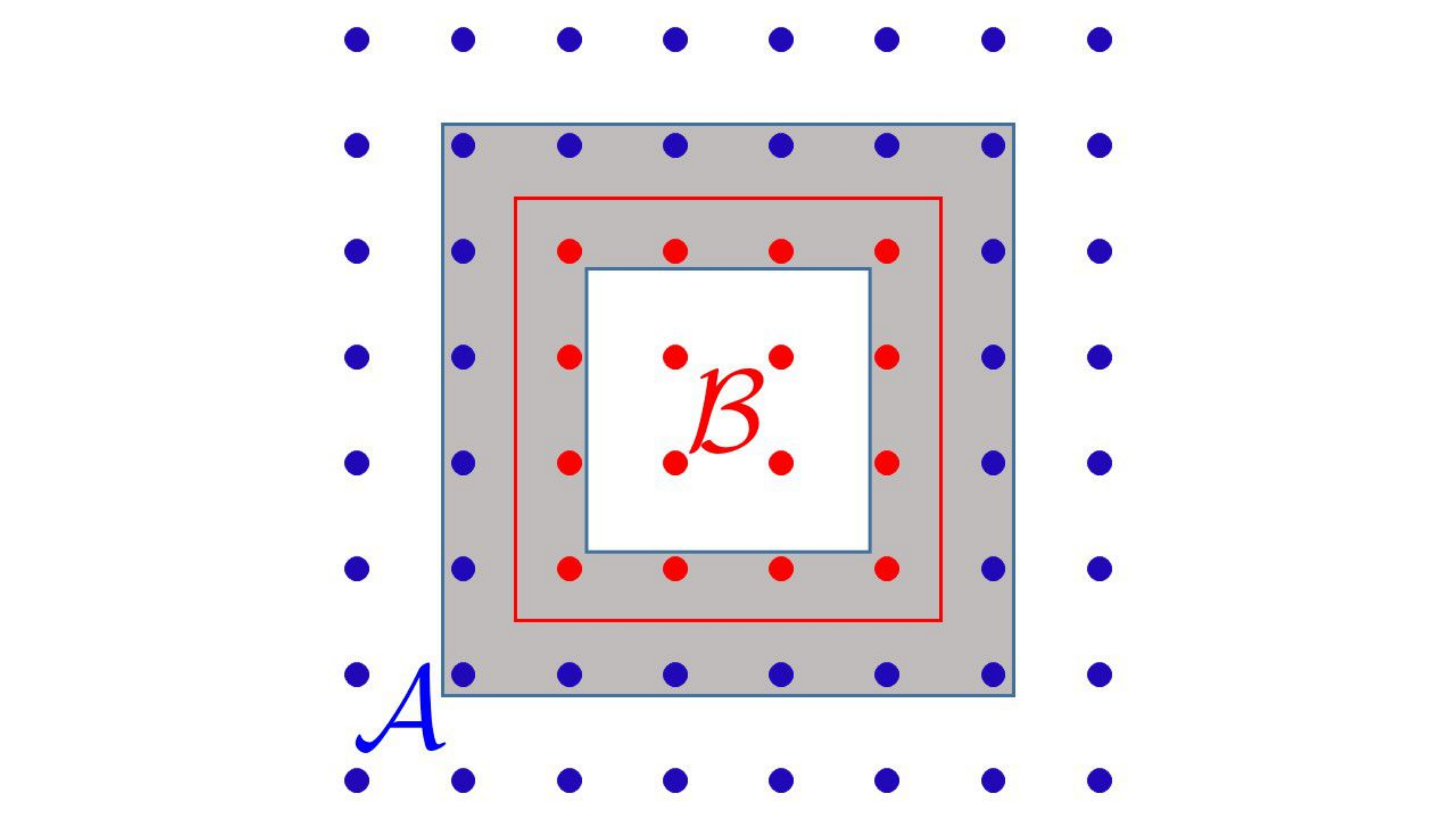} 
   \caption{A 2-D lattice divided by the red contour of length $\ell$ into two disjoint regions ${\cal A}$ and ${\cal B}$. The thickness of the shaded area is the correlation length $\sim \zeta$. The shaded region is the communication channel between ${\cal A}$ and ${\cal B}$. The entanglement entropy and mutual information scale with $\ell$.}
	\label{2D lattice}
\end{figure}

 The area law scaling in noncritical systems is attributed to the fact that there is a finite correlation length $\zeta$ between two disjoint systems ${\cal A}$ and ${\cal B}$. Therefore, regions that are separated by more than $\zeta$ will not contribute to the entanglement entropy \cite{PhysRevLett.100.070502}. To fix ideas, we take a $2$-D lattice and divide it into two disjoint regions  ${\cal A}$ and ${\cal B}$ such that ${\cal A}$ is the complement of ${\cal B}$ and $\ell$ is the length of the boundary between them, see Figure \ref{2D lattice}.  Then, the entanglement entropy takes the general form \footnote{Again, we are neglecting a volume term, which appears at finite temperature; see the above footnote. Entanglement entropy will also have UV divergences in the continuum description, which are cured by putting the system on a lattice. Mutual information, on the other hand, is free from UV divergences.}
\begin{eqnarray}
S({\cal A})={\cal C}\ell+\gamma\,.
\end{eqnarray}
In general, ${\cal C}$ depends on the correlation length $\zeta$, while the constant term $\gamma$ is known as the topological entanglement entropy. Interestingly enough, even in systems that exhibit divergent correlation functions, for example, near criticality, the area law scaling can still be proven to hold\footnote{Entanglement entropy can also have a sub-leading logarithm, $\log \ell$, which is typical in quantum critical systems \cite{Metlitski:2011pr}.} \cite{PhysRevA.73.012309}. This will be the case in the XY-models we study in this work.  Since  mutual information $I(X;Y)$ is the sum of entanglement entropies, it will also follow the area law\footnote{The definition of mutual information, as given by (\ref{Renyi MI}), guarantees that the leading term in $I(X;Y)$ is the area law term, even at finite temperature. In other words, the volume term, which is present in the entanglement entropy at finite temperatures, cancels out in the definition of mutual information.}. However, unlike entropy, which measures the uncertainty about the system, mutual information will quantify the amount of information shared between parts of the system, and hence, it can be a useful tool to detect subtle properties of different phases.  In this work we use both entanglement entropy and mutual information to study the nature of the deconfinement phase transition in dYM, dYM(F), and QCD(adj). 

%%%%%%%%%%%%%%%%%%%%%%%%%%%%%%%%%%%%%%%%%%%%%%%%%%%%%%%%%%%%%%%%%%%%%%
%%%%%%%%%%%%%%%%%%%%%%%%%%%%%%%%%%%%%%%%%%%%%%%%%%%%%%%%%%%%%%%%%%%%%%%%%%%%

%%%%%%%%%%%%%%%%%%%%%%%%%%%%%%%%%
\subsection{The replica trick}
%%%%%%%%%%%%%%%%%%%%%%%%%%%%%%%%%

The calculations of the entanglement entropy and mutual information is notoriously difficult and analytical expressions of these quantities can be obtained only in a few cases. The standard method to calculate the entanglement entropy of a quantum field/statistical field theory is the replica trick: we consider $n$ replicas of the original system and take the limit $n\rightarrow 1$. In order to elucidate the procedure, we start from the generalized R\'enyi entropy defined in (\ref{generalized Reny entropy}) and consider $n=2$. Here, we follow the discussion in \cite{PhysRevB.87.195134}. The joint probability $p(x,y)$ is given by the Boltzmann distribution $p(x,y)=e^{-\beta E(x,y)}/Z$, where $Z=\sum_{x \in X, y\in Y}e^{-\beta E(x,y)}$ and $E(x,y)$ is the energy associated with the states $x \in X$ and $y \in Y$.  The probability $p(x)$ is obtained by tracing over $y$: $p(x)=\sum_{y \in Y}e^{-\beta E(x,y)}/Z$.  Then, the second power of the probability is given by $p^2(x)=\left(\sum_{y \in Y}e^{-\beta E(x,y)}\right)\left(\sum_{y' \in Y}e^{-\beta E(x,y')}\right)/Z^2$. Now, we define the replicated partition function as:
\begin{eqnarray}
Z[{\cal A},2]\equiv \sum_{x \in X} \sum_{y,y' \in Y}e^{-\beta\left(E(x,y)+E(x,y')\right)}\,.
\label{replicated Z}
\end{eqnarray}
Then, the generalized R\'enyi entropy is given from (\ref{reduced generalized  Reny entropy}) as
\begin{eqnarray}
S_2({\cal A})=-\log Z[{\cal A},2]+2\log Z\,.
\label{Reny entropy 2}
\end{eqnarray}
As we discuss below, the replicated partition function (\ref{replicated Z}) can be readily simulated by means of Monte Carlo methods. 

One can easily generalize this discussion to a generic value of $n$ to find that the entanglement entropy is given by the limit
\begin{eqnarray}
S({\cal A})=\lim_{n \rightarrow 1}\frac{1}{1-n}\log\left(\frac{Z[{\cal A},n]}{Z^n}\right)\,.
\end{eqnarray}
The replicated partition function $Z[{\cal A},n]$ is the Boltzmann-weighted sum of fields in ${\cal A}$ and  $n$ replicated (sheets) of fields in ${\cal B}$. Having $n$ replicas is equivalent to formulating the theory on a flat cone with a deficit angle $\delta=2\pi(1-n)$, see \cite{Callan:1994py}.  In a lattice formulation we use a specific number of replicas (in this work we limit our study to $n=2$), while in the continuum it is usually easier to compute the partition function on a cone with an infinitesimal deficit angle.

%%%%%%%%%%%%%%%%%%%%%
\subsection{Overview and strategy for calculating entanglement entropy and mutual information in Yang-Mills on $\mathbb R^2 \times \mathbb T^2$}
\label{Strategy}
%%%%%%%%%%%%%%%%%%%%%

As we mentioned in the introduction, the calculations of entanglement entropy in 4-D confining gauge theories suffer from difficulties due  to strong coupling and nonfactorizability of the gauge invariant Hilbert space on a lattice.  Compactifying the theory on a small circle  results in breaking the gauge group to its $U(1)$ part, and therefore, the 3D spectrum contains a massless photon and a tower of heavy excitations.  Deep in the IR the heavy excitations decouple along with their Hilbert space; they only leave a trace as a renormalization of the $3$-D effective  coupling constant $g_3\equiv g_4/\sqrt{L}$. Since the $3$-D $U(1)$ gauge theory is dual to a compact scalar, its Hilbert space shouldn't suffer from nonfactorizability\footnote{This is specially true when we put the theory on a lattice.}. The compact nature of $U(1)$ allows for magnetic instantons to populate the vacuum giving rise to the confinement phenomenon. When we consider the system at finite temperature (now we are compactifying the time direction) heavy excitations are reintroduced into the partition function of the $3$-D theory via the Boltzmann weight, see Figure \ref{Coulomb gas}. However, only the lightest charged excitations, under $U(1)$, will participate in the dynamics that lead to the deconfinement transition.  Computing the entanglement entropy or mutual information of the 3-D system is a cumbersome task; we don't attempt to do that here. Near the deconfinement transition, however, we can neglect all the heavy Matsubara modes keeping only the zero mode; the system effectively lives in 2D. At this stage we have a 2-D Coulomb gas that is mapped to the XY-spin model with $\mathbb Z_p$ symmetry-preserving perturbations or dual Sine-Gordon model. 

The entanglement entropy/mutual information of these systems will be studied and we shall draw conclusions about the behavior of such information theoretic quantities near deconfinement. Here, one might wonder what happens to the gauge variables in the original gauge theory as well as the ambiguity related to entanglement entropy. The answer is that the gauge theory/ XY-spin model (or dual sine-Gordon model) duality that we use in this work offers escape from the original problem. In a duality there are two equivalent facets of reality, and usually it is much easier to compute certain physical quantities in one picture than the other\footnote{This is very similar to AdS/CFT duality: it is always much easier to compute in one side (usually the AdS side) than the other.}. Thus, woking in the XY-spin (or dual Sine-Gordon) model side of the duality captures all the physical information of the gauge theory (near deconfinement) without having to work with the original gauge variables.

 We start our analysis with the dual Sine-Gordon model and approximate it as a CFT with deformations. This allows us to use the entanglement entropy of CFT to study our system near transition. Next, we calculate R\'enyi mutual information (RMI) of the XY-spin model and study its behavior near the transition. 

In fact, information about the CFT universality class can also be extracted from RMI of the XY-spin model at the critical temperature (assuming that the temperature has been determined precisely). This can be done by computing RMI for different partitions of a given lattice size and trying to fit the next to leading term of RMI (the leading term is being the area term) with general known behavior of CFT at finite interval \cite{PhysRevLett.112.127204}.  An alternative method would be determining the central charge off-criticality by computing the correlators from Monte Carlo simulations. This can give a link between the analytical expressions we obtain for the entanglement entropy in the dual Sine-Gordon model and the numerical computation of RMI in the XY-spin models.  We don't try these calculations in the present work leaving them for a future investigation. 

%%%%%%%%%%%%%%%%%%%%%%%%%%%%%%%%%%%%%%%%%%%%%%%%%%%%%%%%%%%%%%%%%%%%
\subsection{Entanglement entropy in the continuum description: the Sine-Gordon model}
%%%%%%%%%%%%%%%%%%%%%%%%%%%%%%%%%%%%%%%%%%%%%%%%%%%%%%%%%%%%%%%%%%%%

At this stage, we are equipped with enough tools to study mutual information/entanglement entropy of the dual Coulomb gas. Our main purpose is to investigate the interplay between the existence/absence of order parameters and information-theoretic techniques. 

Before starting our systematic study of entanglement entropy, we pause here to discuss the expected behavior of this quantity in each of the theories at hand. At temperatures much lower than the critical temperature, $ T_c$, the system is dominated by magnetic charges. The system is in a gapped phase and information cannot be communicated between distant regions in the plasma. At temperatures much higher than $T_c$ the system is populated by electric charges and again is in a gapped phase. Similar to the magnetic phase, the electric phase does not permit the communication of information over large distances. In addition, there is a region of temperatures in between, where both electric and magnetic charges proliferate. In both dYM and QCD(adj) there is also a point, $T_c$, where the system experiences a phase transition and  develops a massless mode. It is exactly at $T_c$ where one expects to see an inflection point in entanglement entropy, which signals a change in the role played by electric and magnetic components.

%%%%%%%%%%%%%%%%%%%%%%%%%%%%%%%%%%%%%%%%%%%%%%%%%%%%%%%%%%%%%%%%%%%%%%
\begin{figure}[t] %  figure placement: here, top, bottom, or page
   \centering
   \includegraphics[width=6in]{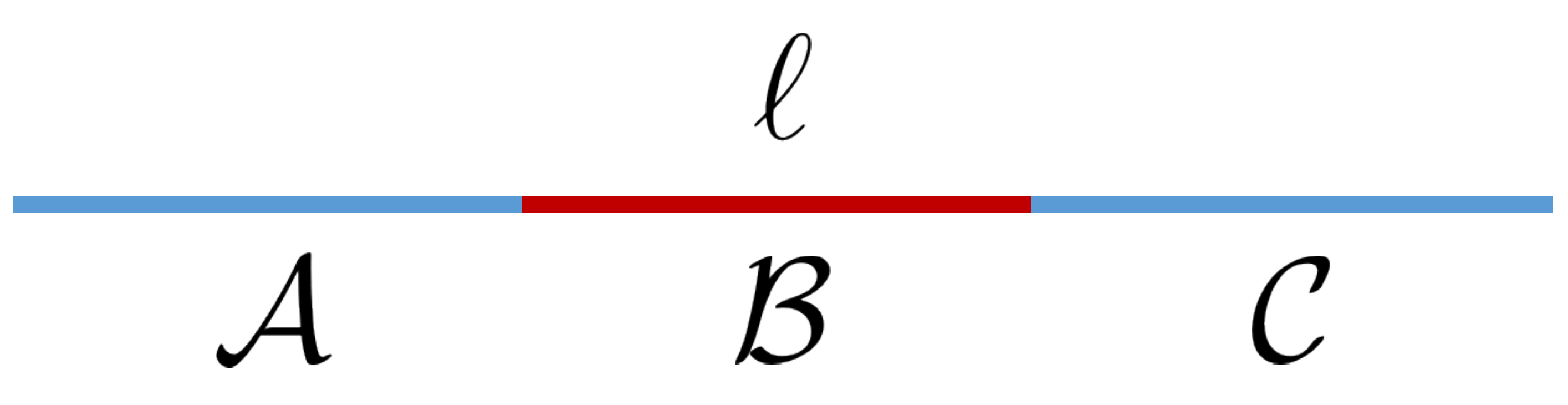} 
   \caption{The one-dimensional space in $1+1$ CFT is tripartitioned into regions ${\cal A}$, ${\cal B}$, and ${\cal C}$. The length of region ${\cal B}$ is $\ell$. }
	\label{figure CFT}
\end{figure}
%%%%%%%%%%%%%%%%%%%%%%%%%%%%%%%%%%%%%%%%%%%%%%%%%%%%%%%%%%%%%%%%%%%%%%%%%%%%

Analytical calculations of the entanglement entropy of the continuum XY-spin model (\ref{general S XY spin model}) is not a straightforward task because of the compact nature of the scalar field. Instead, it is more appropriate to consider the entanglement entropy of the dual Sine-Gordon model. The calculations here are also cumbersome and one needs to find an approximation technique that will enable us to shed light on the entanglement entropy near the transition. 

 As we discussed in Section (\ref{The dual Sine-Gordon model and deconfinement}), there are temperature windows where we can integrate out either the magnetic or electric charges and obtain effective Sine-Gordon models given by (\ref{magnetic sin Gordon}) and (\ref{electric sin Gordon}) for the magnetic and electric disordered phases, respectively. The calculations of the entanglement entropy of the Sine-Gordon model was done in \cite{Banerjee:2016yaf} via perturbation analysis, which treated the model as a free $1+1$D CFT deformed by a primary operator of dimension $\Delta_{\alpha}$ or $\Delta_{\beta}$. The calculations of the entanglement entropy of a CFT demands the partition of space into three regions: ${\cal M}={\cal A}\cup {\cal B}\cup {\cal C}$. This is necessary since a CFT does not have a length scale and one needs to introduce some scale into the problem. In particular, we take the intermediate region ${\cal B}$ to have a length $\ell$, see Figure \ref{figure CFT}. The entanglement entropy of the free CFT is $S_0=\frac{1}{3}\log \frac{\ell}{a}$, where $a$ is a UV cutoff \cite{Calabrese:2004eu}. The change of the entanglement entropy due to a primary operator is then given by \cite{Banerjee:2016yaf}:
\begin{eqnarray}
\nonumber
\Delta S_{\beta}&=&\frac{\beta^2(\mu)}{128}\left(\frac{\rho^2}{4\pi}-2 \right)\log\left(\frac{\ell}{a}\right)\,,\quad \mbox{for magnetically disordered phase}\,,\\
\Delta S_{\alpha}&=&\frac{\alpha^2(\mu)}{128}\left(\frac{\kappa^2}{4\pi}-2 \right)\log\left(\frac{\ell}{a}\right)\,,\quad \mbox{for electrically disordered phase}\,.
\label{change of EE}
\end{eqnarray}
 These expressions are obtained in a regime where perturbation theory is valid, i.e. $\Delta_\alpha>2$ and $\Delta_\beta>2$. In the following we make use of (\ref{change of EE}) to study the behavior of the entanglement entropy of the dual Coulomb gas near the transition temperature. 

%%%%%%%%%%%%%%%%%%%%%%%%%%%%%%%%%%%%%%%%%%%%%%%%%%%%%%%%%%%%%%%%%%%%%
\subsubsection*{Purely electric and purely magnetic systems}
%%%%%%%%%%%%%%%%%%%%%%%%%%%%%%%%%%%%%%%%%%%%%%%%%%%%%%%%%%%%%%%%%%%%

In order to appreciate the role of entanglement entropy in detecting a phase transition or crossover, we first study purely electric and purely magnetic systems. Such systems are gases of one type of charges, either magnetic or electric, and they are  described by the Sine-Gordon models  (\ref{magnetic sin Gordon}) or (\ref{electric sin Gordon}).    Both electric and magnetic gases experience a phase transition at $\Delta_{\alpha}=\Delta_{\beta}=2$, i.e., at $T_c=\frac{g^2}{2\pi L}$. In the magnetic gas the conformal dimension changes from $\Delta_\beta<2$ for $T<T_c$ (magnetic disordered phase)  to $\Delta_\beta>2$ for $T>T_c$ (free phase). While in the electric gas things happen in the reversed order: the conformal dimension changes from $\Delta_\alpha>2$ for $T<T_c$ (free phase)  to $\Delta_\alpha<2$ for $T>T_c$ (electric disordered phase).  This is the celebrated Berezinsky-Kosterlitz-Thouless (BKT) phase transition \cite{Berezinsky:1970fr,Kosterlitz:1973xp}. Furthermore, from (\ref{RG equations}) we find
\begin{eqnarray}
\alpha(\mu)=\alpha_0\left(\frac{\mu}{\mu_0}\right)^{\frac{g^2}{4\pi LT}-2}\,,\quad
\beta(\mu)=\beta_0\left(\frac{\mu}{\mu_0}\right)^{\frac{4\pi LT}{g^2}-2}\,,
\label{solutions of RG}
\end{eqnarray}
where $\alpha_0=e^{-\frac{M_W}{T}}$ and $\beta_0=e^{-\frac{4\pi^2}{g^2}}$ are the UV fugacities, and we have neglected pre-exponential coefficients. We take $\mu_0=a^{-1}$ to be the UV cutoff scale and $\mu=\zeta^{-1}$ to be the correlation length of the system in the IR. Then, we substitute (\ref{solutions of RG}) into (\ref{change of EE}) and expend near $\Delta=2$ to find 
\begin{eqnarray}
\Delta S_{\alpha,\beta}\propto (\Delta_{\alpha,\beta}-2)+{\cal O}\left(\left(\Delta_{\alpha,\beta}-2\right)^2\right)\,.
\end{eqnarray}
Therefore, the change in the entanglement entropy is monotonic across the transition: in the magnetic gas $\Delta S_{\beta}$ interpolates between negative values for $T<T_c$ to positive values for $T>T_c$, while in the electric gas $\Delta S_{\alpha}$ interpolates between positive values for $T<T_c$ to negative values for $T>T_c$, see Figure \ref{figure EE}. Whence, the entanglement entropy itself does not experience a sharp change across the transition point in the purely electric or purely magnetic systems. However, in hybrid systems one expects to see an exchange of the magnetic and electric roles at the transition, and hence, a change in the behavior of the entanglement entropy. 

%%%%%%%%%%%%%%%%%%%%%%%%%%%%%%%%
\subsubsection*{dYM}
%%%%%%%%%%%%%%%%%%%%%%%%%%%%%%%%

\begin{figure}[t] %  figure placement: here, top, bottom, or page
   \centering
   \includegraphics[width=2in]{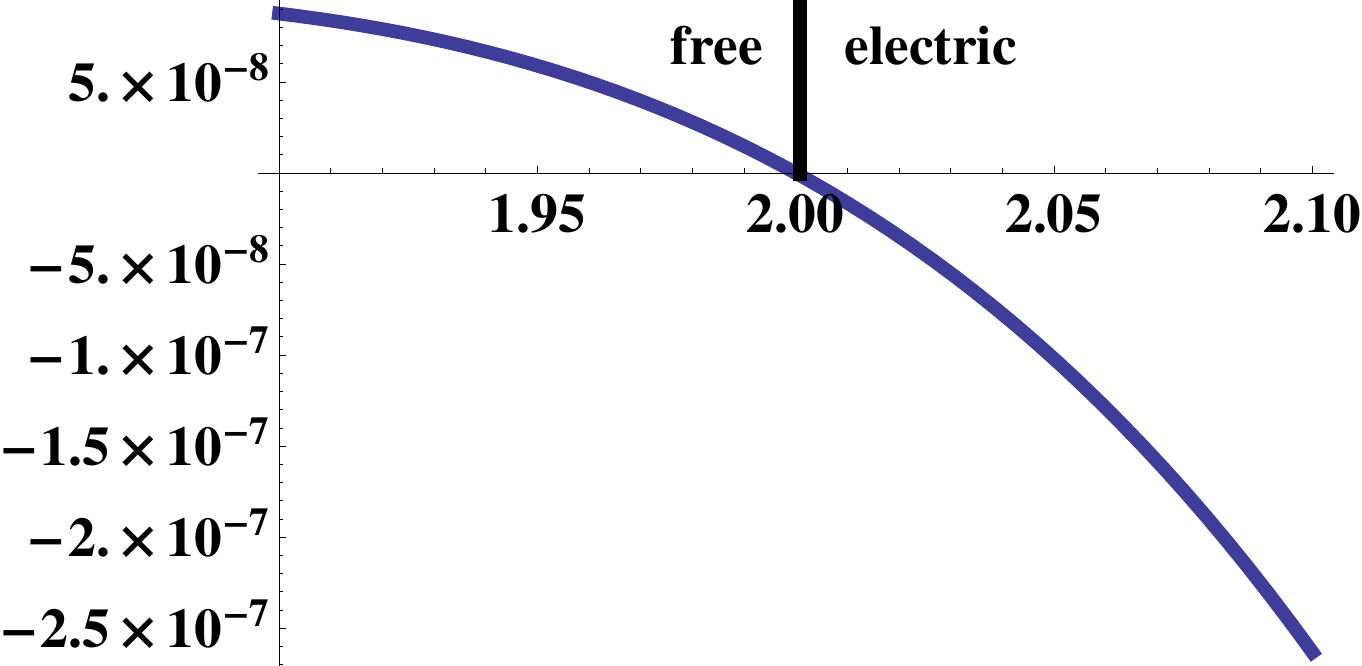} 
   \includegraphics[width=2in]{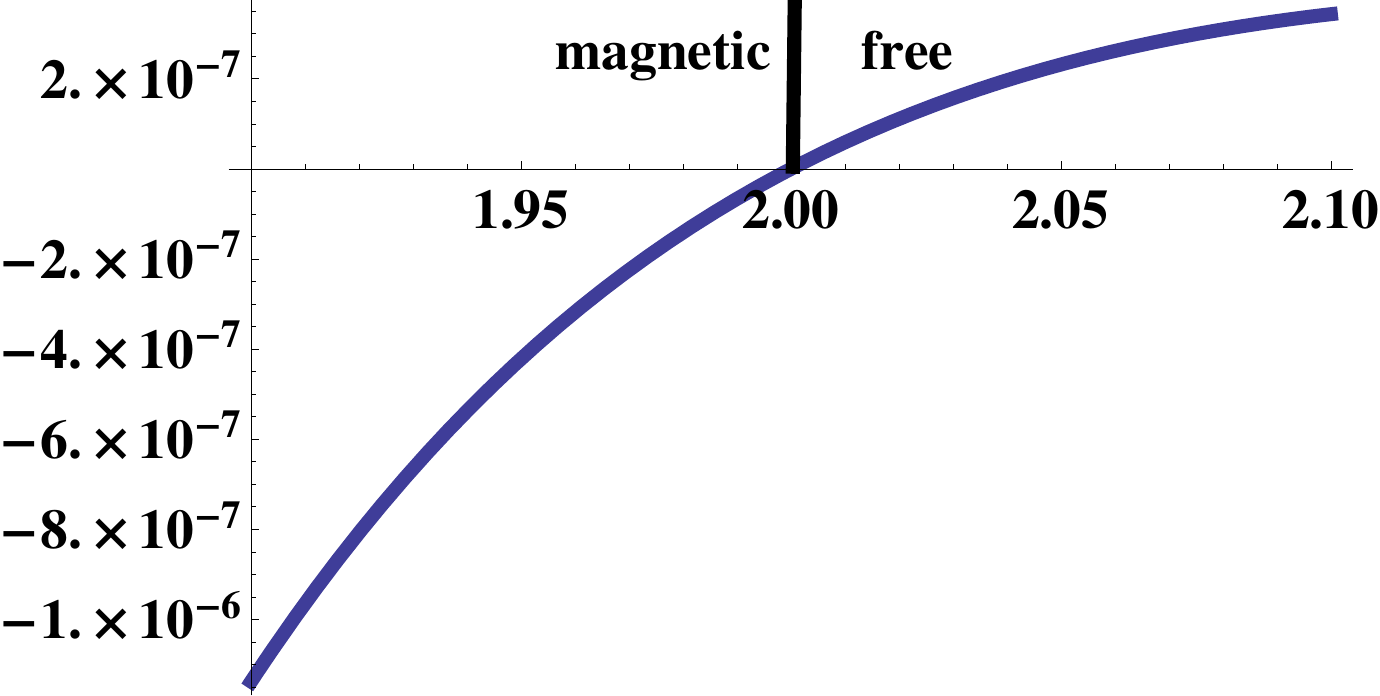}
   \includegraphics[width=2in]{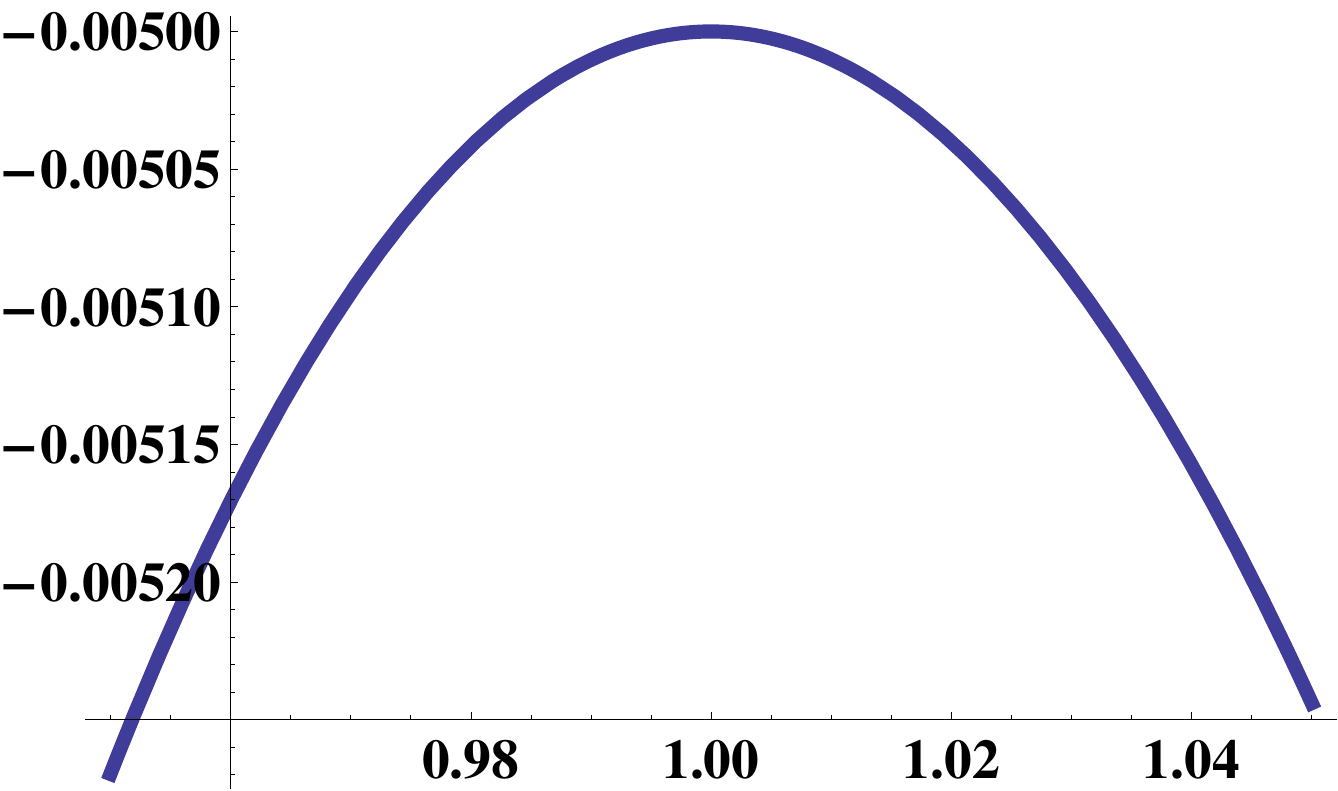}
   \caption{From left to right: the behavior of $\Delta S$ as a function of $x\equiv \frac{4\pi LT}{g^2}$ for the pure electric, pure magnetic, and  dYM Coulomb gases. We use appropriate values of $\zeta$ and $a$ in order to produce the numerical graphs such that $\zeta\gg a$. }
	\label{figure EE}
\end{figure}

The phase transition occurs in the temperature window $\frac{g^2}{8\pi L}<T<\frac{g^2}{2\pi L}$. In this window both electric and magnetic perturbations are relevant ($\Delta_\alpha<2$ and $\Delta_\beta<2$): the theory is strongly coupled and strictly speaking one should not trust (\ref{change of EE}). Nevertheless, one can add both the electric and magnetic contributions to $\Delta S$ in order to crudely study the qualitative behavior of the change of the entanglement entropy near the transition temperature.   Substituting (\ref{solutions of RG}) into (\ref{change of EE}) and assuming that $\zeta\gg \ell\gg a$, we find the total change of the entanglement entropy
\begin{eqnarray}
\nonumber
\Delta S_{dYM}&=&\Delta S_\alpha+\Delta S_\beta\\
\nonumber
&=&\frac{\log\left(\frac{\ell}{a}\right)}{128}\left\{\alpha_0^2\left(\frac{g^2}{4\pi LT}-2\right)\left(\frac{a}{\zeta}\right)^{\frac{g^2}{2\pi LT}-4}+ \beta_0^2\left(\frac{4\pi LT}{g^2}-2\right)\left(\frac{a}{\zeta}\right)^{\frac{8\pi LT}{g^2}-4}\right\}\,,\\
\end{eqnarray}
where $\alpha_0=e^{-\frac{M_W}{T}}$, $\beta_0=e^{-\frac{4\pi^2}{g^2}}$. This quantity attains a maximum at $T_{max}=\frac{g^2}{4\pi L}$, see Figure \ref{figure EE},  which is exactly the transition temperature obtained via bosonization. 

%%%%%%%%%%%%%%%%%%%%%%%%%%%%%%%%%%%%%%%%%%%%%%%%%%%%%%%%%%%%%%%%%%%%%%
%%%%%%%%%%%%%%%%%%%%%%%%%%%%%%%%%%%%%%%%%%%%%%%%%%%%%%%%%%%%%%%%%%%%%%%%%%%%

%%%%%%%%%%%%%%%%%%%%%%%%%%%%%%%%
\subsubsection*{QCD(adj)}
%%%%%%%%%%%%%%%%%%%%%%%%%%%%%%%%

\begin{figure}[t] %  figure placement: here, top, bottom, or page
   \centering
			\includegraphics[width=2in]{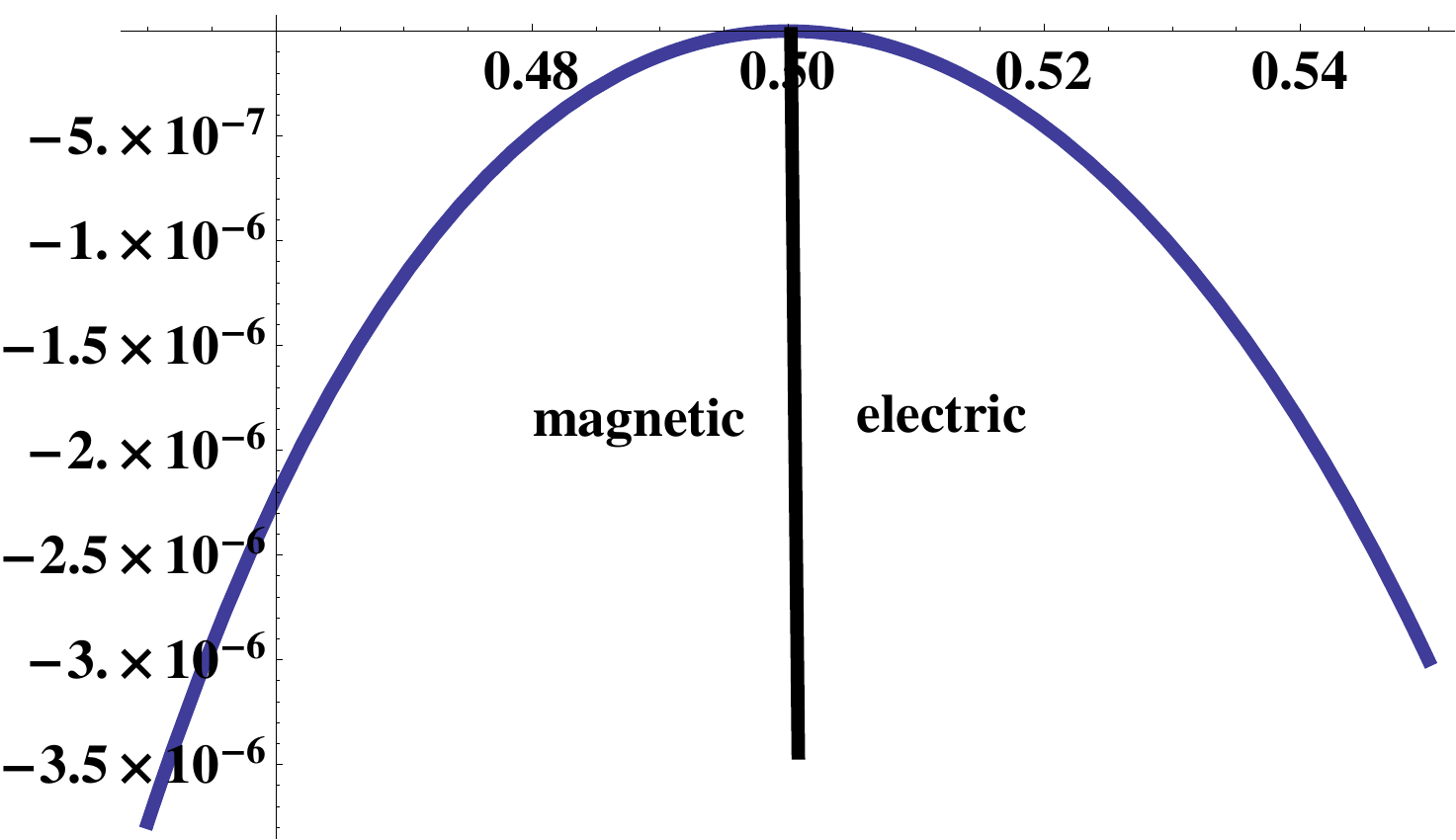} 
			\hspace{2cm}
			\includegraphics[width=2in]{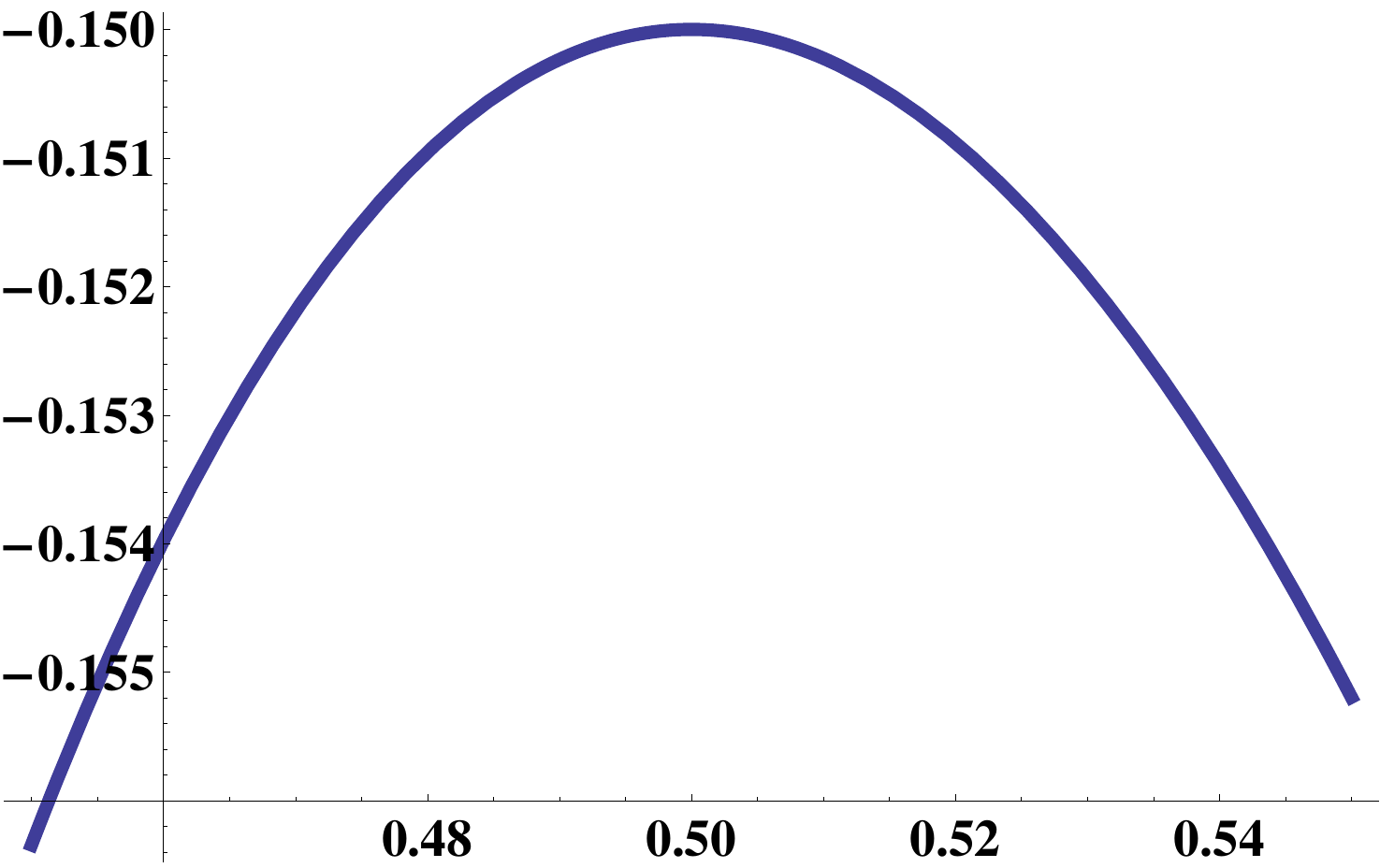}
   \caption{The behavior of $\Delta S$ as a function of $x\equiv \frac{4\pi LT}{g^2}$. Left: $\Delta S$ of QCD(adj). $\Delta S$ attains a maximum at $T_c=\frac{g^2}{8\pi L}$. In addition $\Delta S=0$ exactly at $T_c$, which indicates that the theory is Gaussian at the transition point. Right: $\Delta S$ of dYM(F).}
	\label{figure EE for QCDadj}
\end{figure}

We can repeat the same exercise above for the dual Sine-Gordon model of QCD(adj). The resulting change in entropy is given by:
\begin{eqnarray}
\nonumber
\Delta S_{QCD(adj)}=\frac{\log\left(\frac{\ell}{a}\right)}{128}\left\{\alpha_0^2\left(\frac{g^2}{4\pi LT}-2\right)\left(\frac{a}{\zeta}\right)^{\frac{g^2}{2\pi LT}-4}+ \beta_0^2\left(\frac{16\pi LT}{g^2}-2\right)\left(\frac{a}{\zeta}\right)^{\frac{32\pi LT}{g^2}-4}\right\}\,,\\
\end{eqnarray}
where $\alpha_0=e^{-\frac{M_W}{T}}$, $\beta_0=e^{-\frac{8\pi^2}{g^2}}$. The change in the entanglement entropy has a maximum at $T_{max}=\frac{g^2}{8\pi L}$, which is again the critical temperature. Interestingly enough, we find $\Delta S_{QCD(adj)}(T=T_{max}=T_c)=0$, see Figure \ref{figure EE for QCDadj}. This shows that the entanglement entropy does not get any additional contribution at $T_c$. Hence, the theory is free at $T_c$, the same conclusion that can be reached via more advanced CFT technology. 

%%%%%%%%%%%%%%%%%%%%%%%%%%%%%%%%%%%%%%%%%%%%%%%%%%%%%%%%%%%%%%%%%%%%%%
%%%%%%%%%%%%%%%%%%%%%%%%%%%%%%%%%%%%%%%%%%%%%%%%%%%%%%%%%%%%%%%%%%%%%%%%%%%%

%%%%%%%%%%%%%%%%%%%%%%%%%%%%%%%%
\subsubsection*{dYM(F)}
%%%%%%%%%%%%%%%%%%%%%%%%%%%%%%%%

Now, let us consider the same quantity for dYM(F):
\begin{eqnarray}
\nonumber
\Delta S_{dYM(F)}=\frac{\log\left(\frac{\ell}{a}\right)}{128}\left\{\alpha_0^2\left(\frac{g^2}{16\pi LT}-2\right)\left(\frac{a}{\zeta}\right)^{\frac{g^2}{8\pi LT}-4}+ \beta_0^2\left(\frac{4\pi LT}{g^2}-2\right)\left(\frac{a}{\zeta}\right)^{\frac{8\pi LT}{g^2}-4}\right\}\,,\\
\end{eqnarray}
where $\alpha_0=e^{-\frac{M_F}{T}}$, $\beta_0=e^{-\frac{4\pi^2}{g^2}}$. Despite the fact that the theory is always in a gapped phase, nevertheless, the change in entanglement entropy has a maximum at $T_{max}=\frac{g^2}{8\pi L}$, see Figure \ref{figure EE for QCDadj}. We anticipate that a cross over happens at this temperature.

%%%%%%%%%%%%%%%%%%%%%%%%%%%%%%%%%%%%%%%%%%%%%%%%%%%%%%%%%%%%%%%%%%%%%%%%%%%%%%%%%%%%%%%%%%%%%%
\subsection{Entanglement entropy and mutual information on the lattice: A Monte Carlo setup}
\label{Entanglement entropy and mutual information on the lattice: A Monte Carlo setup}
%%%%%%%%%%%%%%%%%%%%%%%%%%%%%%%%%%%%%%%%%%%%%%%%%%%%%%%%%%%%%%%%%%%%%%%%%%%%%%%%%%%%%%%%%%%%%%

%%%%%%%%%%%%%%%%%%%%%%%%%%%%%%%%%%%%%%%%%%%%%%%%%%%%%%%%%%%%%%%%%%%%%%
\begin{figure}[t] %  figure placement: here, top, bottom, or page
   \centering
			\includegraphics[width=2.5in]{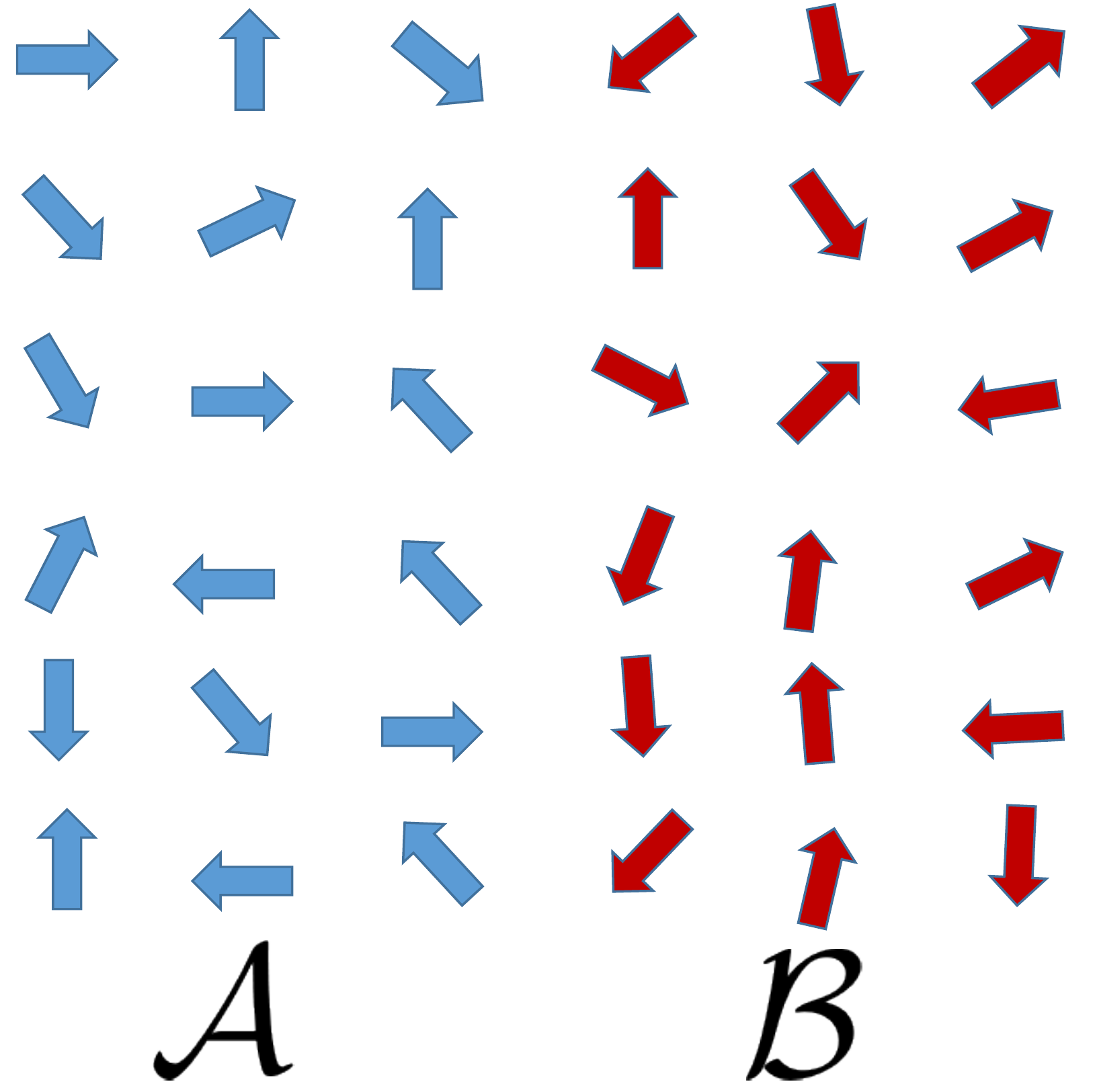}
			\hspace{1cm}
			\includegraphics[width=2.5in]{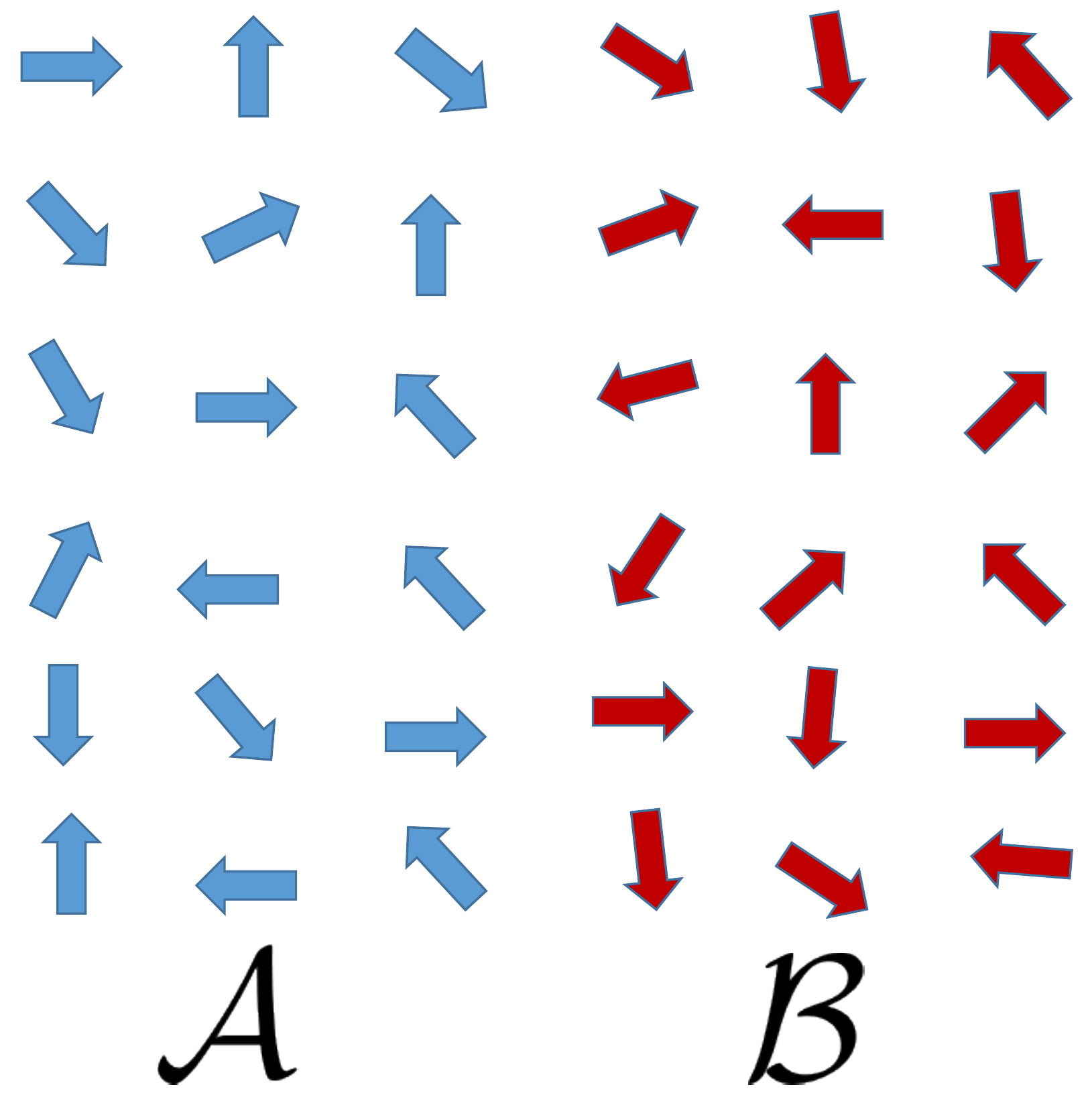}
   \caption{A typical configuration in calculating the second R\'enyi mutual information. There are two replicas (left and right) and each one is divided into two regions ${\cal A}$ and ${\cal B}$. Regions ${\cal A}$ of the two replicas are strongly correlated in the sense that an update of any spin in ${\cal A}$ is accepted only and only if the update affects the same spin in both replicas. On the other hand, the updates in regions ${\cal B}$ are independent in each replica. }
	\label{replica method on the lattice}
\end{figure}
%%%%%%%%%%%%%%%%%%%%%%%%%%%%%%%%%%%%%%%%%%%%%%%%%%%%%%%%%%%%%%%%%%%%%%%%%%%%

The replica method enables us to compute the entanglement entropy and mutual information on the lattice \cite{PhysRevLett.106.135701,PhysRevLett.104.157201}. As we stressed before, unlike the entanglement entropy, which tells us about the amount of uncertainty in a system, mutual information, $I(X;Y)$, quantifies the amount of information shared between different parts of the system. Fortunately enough, one can calculate  $I(X;Y)$ on the lattice using Monte Carlo methods. Here, we focus on the second R\'enyi Mutual information $I_2(X;Y)$ and consider the situation of a collection of spins located at the lattice sites. To this end, we bipartition a lattice ${\cal M}$ into two regions ${\cal A}$ and ${\cal B}$ and consider two replicas ${\cal M}_1$ and ${\cal M}_2$ such that ${\cal M}_1={\cal A}_1 \cup {\cal B}_1$ and ${\cal M}_2={\cal A}_2 \cup {\cal B}_2$. Now, we apply a boundary condition on the regions such that for a given configuration of spins on ${\cal A}_1$ and ${\cal A}_2$, which is taken to be the exact same configuration in both ${\cal A}_1$ and ${\cal A}_2$, we allow the spins in ${\cal B}_1$ and ${\cal B}_2$ to fluctuate independently, see Figure \ref{replica method on the lattice}. This boundary condition amounts to tracing over the states of system ${\cal B}$ for a given state in ${\cal A}$. The partition function of the system is then given by the replicated partition function  (\ref{replicated Z}). According to $Z[{\cal A},2]$,  Monte Carlo simulations will use the energy $E(x,y)+E(x,y')$ to update the spin moves, which cannot be accepted unless it satisfies the above mentioned boundary condition.  To be more specific, let us consider the Hamiltonian and partition function
\begin{eqnarray}
E=-\sum_{\langle I,J\rangle}\bm S_I \cdot \bm S_J\,,\quad Z=\sum_{\{S_I\}}e^{-E/T}\,,
\end{eqnarray}
where the bracket indicates a sum over nearest neighbor pairs of spins. Then, the total  energy of the replicated system is given by
\begin{eqnarray}
E(x,y)+E(x,y')=-2\sum_{\langle I_A, I_A'\rangle} \bm S_{I_A}\cdot \bm S_{I_A'}- \sum_{\langle I_B, I_B'\rangle} \bm S_{I_B}\cdot \bm S_{I_B'}- \sum_{\langle J_B, J_B'\rangle} \bm S_{J_B}\cdot \bm S_{J_B'}\,.
\end{eqnarray}
We see  that there is a factor of $2$ multiplying the first sum, which indicates  that the effective temperature of region $\cal{A}$ is $T/2$. Hence,  one needs to distinguish between three temperature ranges in the replicated system:
\begin{enumerate}
\item $0<T<T_c$, where $T_c$ is the critical temperature of the non-replicated system: both regions $\cal {A}$ and $\cal {B}$ are below criticality. 
\item $T_c<T<2T_c$: region $\cal {B}$ is above criticality, while region ${\cal A}$ is below it.
\item $T>2T_c$: both regions $\cal {A}$ and $\cal{B}$ are above criticality.
\end{enumerate} 

Monte Carlo simulations don't allow the direct computation of the partition function or entropy. In order to extract the entropy from simulations, one needs to integrate the energy estimator over a range of temperatures. The expectation value of energy is given by $
\langle E \rangle =-\frac{\partial \log Z}{\partial \beta}$, and hence, using the definition (\ref{Reny entropy 2})  we find
\begin{eqnarray}
S_2({\cal A};T)=\int_T^\infty \frac{dT'}{T'^2}\left[\langle E \rangle_{\cal A}(T') -2\langle E\rangle_0(T')\right]\,, 
\label{Reny entropy 2 from integral}
\end{eqnarray}
where $\langle E \rangle_{\cal A}$ and $\langle E\rangle_0$ are respectively the energy expectation values of the replicated and original (non-replicated) systems. Similarly, the R\'enyi Mutual information is given by the expression
\begin{eqnarray}
I_2(X;Y;T)=\int_T^\infty \frac{dT'}{T'^2}\left[2\langle E \rangle_{\cal A}(T') -2\langle E\rangle_0(T')-\langle E\rangle_{{\cal A}\cup {\cal B}}(T')\right]\,, 
\label{mutual info 2 from integral}
\end{eqnarray}
where $\langle E\rangle_{{\cal A}\cup {\cal B}}(T)$ is the energy of the replicated system as we shrink ${\cal B}$ to $\emptyset$, i.e., it is the energy of the original system at $T/2$.
 
In practice, we cutoff the integrals (\ref{Reny entropy 2 from integral}) and (\ref{mutual info 2 from integral}) at some $T_{max}\gg T$. Therefore, the extraction of entanglement entropy or mutual information in Monte Carlo method requires simulations over a large range of temperatures, an expensive and long process. Below, we show how one can partially circumvent this difficulty by making use of the T-dual description of the XY-spin lattice, which also eliminates unwanted vortices with lower winding number.

%%%%%%%%%%%%%%%%%%%%%%%%%%%%%%%%%%%%%%%%%%%%%%%%%%%%%%%%%%%%%%%%%%%%%%%%%%%%%%%%%%%%%%%%
\subsection{Mutual information from XY-spin models on the lattice and T-duality} 
\label{Mutual information from the XY-spin model on the lattice T duality}   
%%%%%%%%%%%%%%%%%%%%%%%%%%%%%%%%%%%%%%%%%%%%%%%%%%%%%%%%%%%%%%%%%%%%%%%%%%%%%%%%%%%%%%%%

As we showed above, the use of information theoretic techniques demands that we partition the system into two or more disjoint regions. This procedure introduces ambiguities in the lattice gauge theory calculations. Fortunately enough, we found that the gauge theory upon compactification is dual to  XY-spin models. Such models do not suffer from ambiguities when studied on a lattice, and the extraction of entanglement entropy and mutual information from these systems is a more straightforward task. 

The lattice version of the continuum XY-spin model (\ref{general S XY spin model}) is given by 
\begin{eqnarray}   
E=-\frac{K}{2\pi}\sum_{\langle I,J \rangle}\cos\left(\theta_I-\theta_J\right)-2G_p\sum_{I}\cos\left(p\theta_I\right)\,,\quad Z=\int_{0}^{2\pi}\prod_i d\theta_i e^{-E}
\label{lattice XY}
\end{eqnarray}
where we set the lattice spacing $a=1$. The equivalence between (\ref{lattice XY}) and  (\ref{general S XY spin model}) is easily shown by expanding the first term in (\ref{lattice XY}) to second order and taking $a \rightarrow 0$. As we showed in Section (\ref{Equivalence between dual Coulomb gas and XY-spin model}), there exists two equivalent XY-spin models for each of the theories we consider in this work. These models are the T-dual of eachother. This conclusion applies also to the lattice formulation, as we discuss momentarily. To be more specific we take QCD(adj) as an example. dYM and dYM(F) follow the same pattern. 

In one of the descriptions the lattice partition function of QCD(adj) is given by (this is the lattice version of the continuum description  (\ref{XY for QCDadj})) 
\begin{eqnarray}
\nonumber
&&E=-\frac{g^2}{16\pi^2 }\sum_{\langle I,J \rangle}\cos\left(\theta_I-\theta_J\right)-2e^{-\frac{8\pi^2}{g^2}} \sum_{I}\cos\left(2\theta_I\right)\,,\\
&&Z[K=\frac{g^2}{8\pi T},G_2=\xi_B,p=2;H_2,w=2]=\int_{0}^{2\pi}\prod_i d\theta_i e^{-E/T}\,,
\label{lattice XY for QCDadj}
\end{eqnarray}
and we have set the size of the $\mathbb S_L^1$ circle equal to the lattice spacing, i.e., $L=a=1$. The description (\ref{lattice XY for QCDadj}) has two pitfalls. First, one needs to strict the monodromies of $\{\theta_I\}$ to be even integers multiples of $2\pi$. This is necessary in order to eliminate the unit winding vortices from the spectrum of the theory (votices of unit windings are fundamental electric charges, which are absent in QCD(adj)). Second, as we found in Section (\ref{Entanglement entropy and mutual information on the lattice: A Monte Carlo setup}), and according to Eq. (\ref{mutual info 2 from integral}), the extraction of mutual information from (\ref{lattice XY for QCDadj}) entails performing extended Monte Carlo simulations. 

%%%%%%%%%%%%%%%%%%%%%%%%%%%%%%%%%%%%%%%%%%%%%%%%%%%
\subsubsection*{The T-dual lattice description}
\label{The T-dual lattice description}
%%%%%%%%%%%%%%%%%%%%%%%%%%%%%%%%%%%%%%%%%%%%%%%%%%%

In order to overcome these drawbacks, we switch to the T-dual description of (\ref{lattice XY for QCDadj}). This is the lattice version of (\ref{dual XY for QCDadj}):
\begin{eqnarray}
\nonumber
&&E= -\frac{4 }{g^2}\sum_{\langle I,J \rangle}\cos\left(\theta_I-\theta_J\right)-2e^{-\frac{M_W}{T}}\sum_{I}\cos\left(4\theta_I\right)\,,\\
&&Z[K=\frac{8\pi T}{g^2},G_4=\xi_W,p=4;H_1,w=1]=\int_{0}^{2\pi}\prod_i d\theta_i e^{-TE}\,.
\label{lattice dual XY for QCDadj}
\end{eqnarray}
Now, we need not worry about suppressing lower-winding vertices in Monte Carlo simulations since magnetic bions (the magnetic excitations of QCD(adj)) in this description have unit windings. The same conclusion can be reached for dYM and dYM(F).

The coefficients that appear in the energy functional (\ref{lattice dual XY for QCDadj}) (or the energy functional of dYM and dYM(F)) are not suitable for realistic Monte Carlo simulations given the extremely small values of the coupling constant and fugacities. Instead of (\ref{lattice dual XY for QCDadj}), we replace it with the phenomenological model:
\begin{eqnarray}
\nonumber
&&E= -\sum_{\langle I,J \rangle}\cos\left(\theta_I-\theta_J\right)-\tilde y\sum_{I}\cos\left(p\theta_I\right)\,,\\
&&Z[\tilde y,p;H_1,w=1]=\int_{0}^{2\pi}\prod_i d\theta_i e^{-TE}\,.
\label{lattice dual XY general}
\end{eqnarray}
This model is capable of capturing the essential features of dYM, dYM(F), and QCD(adj) as follows:
\begin{enumerate}
\item $p=1$. This is dYM(F), where $p=1$  accounts for fundamental quarks and $T\tilde y$ is their fugacity. In principle, one should also add $\sum_{I}\cos(2\theta_I)$ term to account for the W-bosons. However, the W-boson fugacity is exponentially small compared to that of the fundamental quarks, and it is more appropriate to neglect the W-bosons all together in the description.  The unit-winding vortices, $w=1$, are magnetic monopoles. 
\item $p=2$. This is dYM, where $p=2$ accounts for the W-bosons and $T\tilde y$ is their fugacity. Again, the unit-winding vortices are the magnetic monopoles. 
\item $p=4$. This is QCD(adj), where $p=4$ denotes  the W-bosons and $T\tilde y$ is their fugacity. The unit-winding vortices are the magnetic bions. 
\end{enumerate} 
In all cases, exciting a unit-winding vortex costs a core energy, roughly,  ${\cal O}(T)$ in lattice units, which is determined by the kinetic term in (\ref{lattice dual XY general}). Therefore, vortices are suppressed in the high temperature phase. On the other hand, as temperature increases, the fugacity of the electric excitations (fundamental quarks or W-bosons) increases, and hence, their core energies decrease\footnote{The core energy $E_c$ is given by $E_c=-\log\xi$, where $\xi$ is the fugacity.}. Thus, the electric excitations dominate the plasma at high temperatures. This is exactly the expected behavior in dYM, dYM(F), and QCD(adj), which is captured by the phenomenological model (\ref{lattice dual XY general}). 

Although the phenomenological model (\ref{lattice dual XY general}) has ${\cal O}(1)$ fugacities, as opposed to the original system (\ref{lattice dual XY for QCDadj}), which has an exponentially small fugacity owing to its semi-classical nature, it still captures the qualitative features of (\ref{lattice dual XY for QCDadj}) since both models are expected to  belong to the same universality class.  For example, renormalization-group analysis of  (\ref{lattice dual XY for QCDadj}) (XY-pin model with $\mathbb Z_4$ symmetry-preserving perturbations and exponentially small fugacities) showed that it exhibits a continuous phase transition with a fugacity-dependent critical exponent \cite{Anber:2011gn}. This behavior was also confirmed by Monte Carlo simulations of the phenomenological model (\ref{lattice dual XY general}), i.e., for large fugacities, see \cite{PhysRevB.69.174407}. 
 
Now, we come to the point of extracting R\'enyi mutual information from (\ref{lattice dual XY general}). It is trivial to see that the expectation value of energy is $\langle E \rangle=-\frac{\partial Z}{\partial T}$, which replaces the traditional expression $\langle E \rangle=-\frac{\partial Z}{\partial \beta}$. This relation can be inverted to write the logarithm of the partition function as an integral over the energy estimator $\log Z=-\int_0^T dT' \langle E \rangle(T')$. Now, we make use of the definition (\ref{Reny entropy 2}) to find 
\begin{eqnarray}
I_2(X;Y;T)=\int_0^T dT'\left[2\langle E \rangle_{\cal A}(T') -2\langle E\rangle_0(T')-\langle E\rangle_{{\cal A}\cup {\cal B}}(T')\right]\,. 
\label{dual mutual info 2 from integral}
\end{eqnarray}
It is remarkable that the T-dual lattice model (\ref{lattice dual XY general}) provides a neat and cheap method to extract the mutual information compared to the original prescription (\ref{mutual info 2 from integral}), where one needs to suppress lower winding vortices. 

When using the replica method (we use only two replicas in this work) to compute $I_2(X;Y;T)$, one needs to distinguish between three temperature regimes in (\ref{lattice dual XY general}) (as usual we divide our lattice into two regions ${\cal A}$ and ${\cal B}$ such that the spins of regions ${\cal A}$ of the two replicas are updated simultaneously). Since the temperature $T$ multiplies the energy functional in (\ref{lattice dual XY general}), region $A$ will effectively be at temperatures twice that of the original system. The three temperature regimes are:
\begin{enumerate}
\item $0<T<T_c/2$, where $T_c$ is the critical temperature of the non-replicated system: both regions $\cal {A}$ and $\cal {B}$ are below criticality. 
\item $T_c/2<T<T_c$: region $\cal {A}$ is above criticality, while region ${\cal B}$ is below it.
\item $T>T_c$: both regions $\cal {A}$ and $\cal{B}$ are above criticality.
\end{enumerate}

In the next section we perform numerical simulations of (\ref{lattice dual XY general}) and extract lessons from $I_2(X;Y;T)$ about the deconfinement phase transition/crossover. 

%%%%%%%%%%%%%%%%%%%%%%%%%%%%%%%%%%%%%%%%%%
\section{Monte Carlo simulations}
\label{Monte Carlo simulations}
%%%%%%%%%%%%%%%%%%%%%%%%%%%%%%%%%%%%%%%%%%

This section is devoted to the numerical simulations of (\ref{lattice dual XY general}). In particular, we show that mutual information can be used as a probe to detect phase transitions in our theories.  

%%%%%%%%%%%%%%%%%%%%%%%%%%%%%%%%%%%%%%%%%%%%%%%%%%%%%%%%%%%%%%%%%%%%%%%%
\begin{figure}[t] %  figure placement: here, top, bottom, or page
   \centering
			\includegraphics[width=5in]{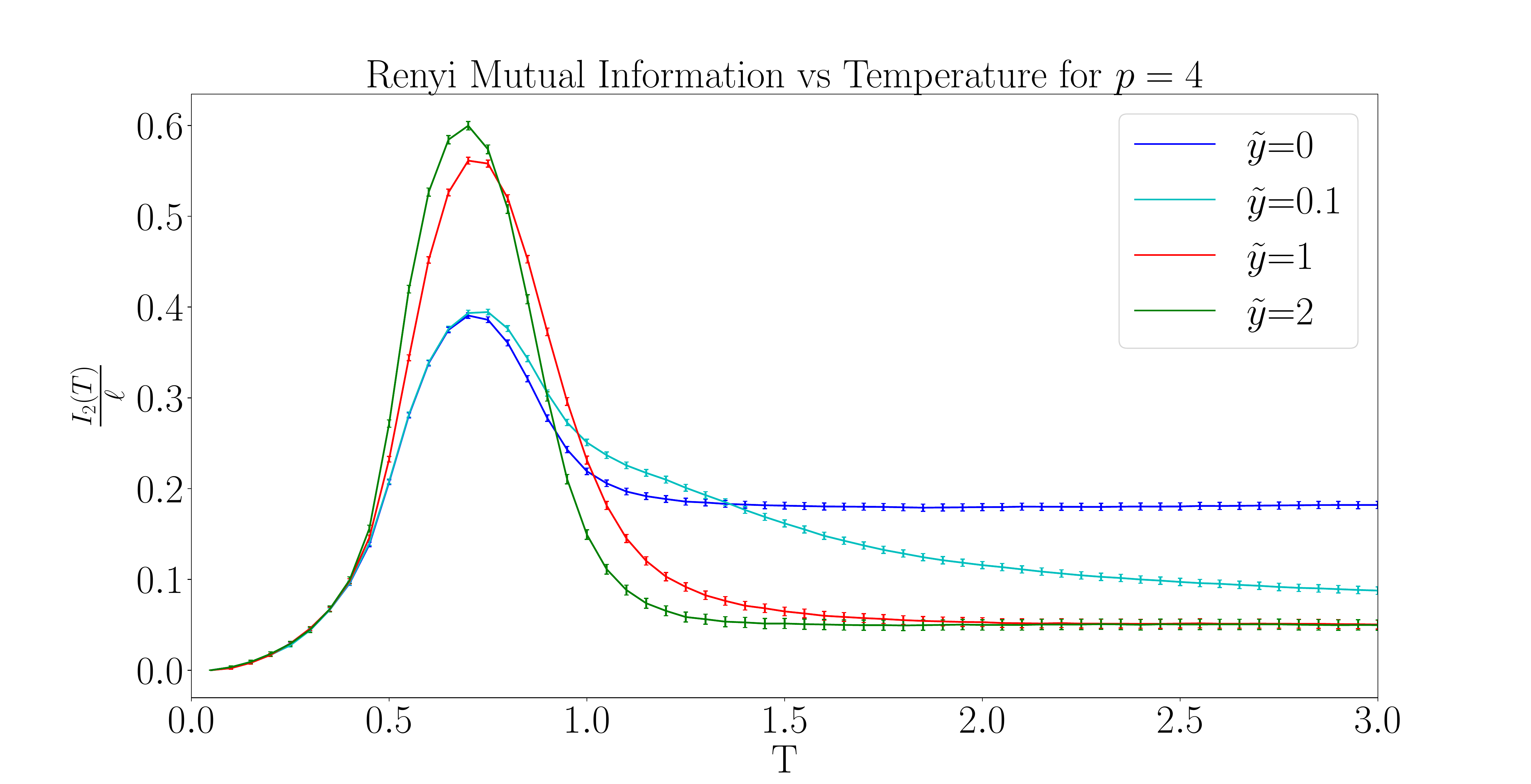} 
   \caption{Mutual information of (\ref{lattice dual XY general}) for $p=4$ and various values of $\tilde y$. We use lattice size $N=16$.}
	\label{different y tilde for N 16}
\end{figure}
%%%%%%%%%%%%%%%%%%%%%%%%%%%%%%%%%%%%%%%%%%%%%%%%%%%%%%%%%%%%%%%%%%%%%%%%%%%%

We use a single-flip Metropolis algorithm and divide our periodic lattice of size $N\times N$ into two regions ${\cal A}={\cal B}$, such that each region is $N\times N/2$ cylinder embedded in $N\times N$ torus. We start by studying R\'enyi mutual information (RMI) of (\ref{lattice dual XY general}) with $p=4$, QCD(adj), and various values of $\tilde y$. The results are shown in Figure \ref{different y tilde for N 16}, where we plot $I_2(X;Y;T)/\ell$  against the temperature and $\ell=2N$ is the length of the boundary between regions ${\cal A}$ and ${\cal B}$. First, we see that all RMI curves coincide at small $T$, irrespective of the value of $\tilde y$. This is consistent with the fact that the electric excitations are confined at low temperatures, their fugacities are irrelevant, and the system is dominated by a plasma of magnetic charges. The correlation length in a plasma is extremely small and the different parts of the system are uncorrelated. This is reflected in the fact that RMI is vanishingly small at low temperature.  As we dial up $T$, the density of the magnetic charges decreases, the correlation length increases, and information can be communicated across larger distances. This can be seen as a spike in RMI, with a magnitude that depends on the value of $\tilde y$.  At high enough temperatures (above the critical temperature $T_c$; we will determine $T_c$ below) RMI asymptotes to a constant value, which decreases with increasing $\tilde y$. In order to understand the significance of this behavior, we compare $\tilde y=0$ with $\tilde y=0.1,1.0,2.0$. The former case corresponds to eliminating W-bosons from our theory. In this case the system exhibits a BKT phase transition, from a massive to massless phase, as we dial up the temperature. This is in contradistinction with the case $\tilde y>0$: dialing up the temperature will cause the system to transit from a massive (magnetic) phase to another massive (electric) phase. Obviously, a massless phase can communicate information more effectively than a massive one, and thus, at high enough temperatures RMI attains larger values. Also, the bigger the value of $\tilde y$, the higher the density of W-bosons in the electric disordered phase and the lower the value of the asymptotic RMI.

%%%%%%%%%%%%%%%%%%%%%%%%%%%%%%%%%%%%%%%%%%%%%%%%%%%%%%%%%%%%%%%%%%%%%%%%
\begin{figure}[t] %  figure placement: here, top, bottom, or page
   \centering
			\includegraphics[width=5in]{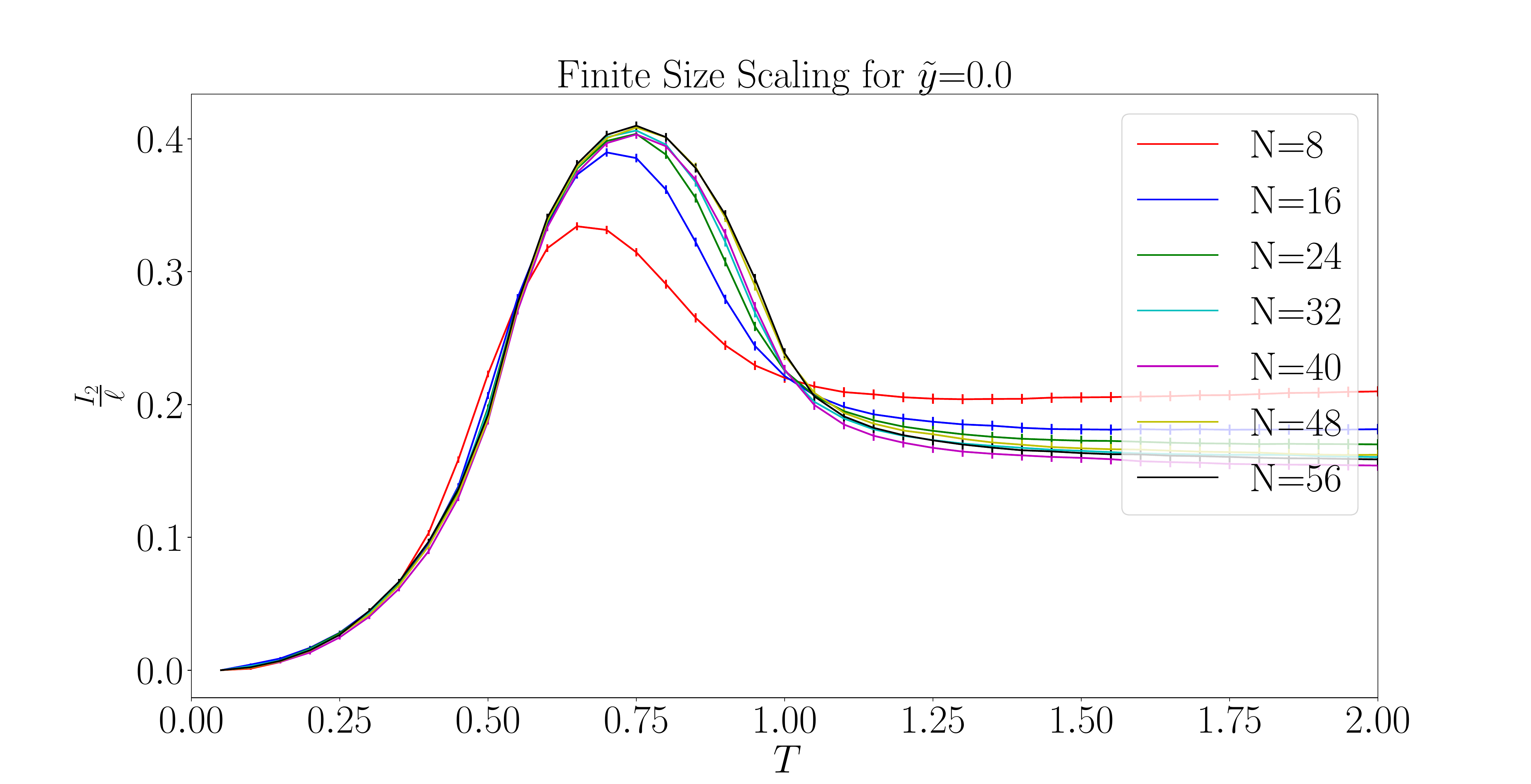} 
   \caption{Finite size scaling for RMI of (\ref{lattice dual XY general}) with $\tilde y=0$. The curves cross at $T\cong 0.5$ and $T\cong1$, which are the values of $T_c/2$ and $T_c$, respectively.}
	\label{finite size scaling y is 0}
\end{figure}
%%%%%%%%%%%%%%%%%%%%%%%%%%%%%%%%%%%%%%%%%%%%%%%%%%%%%%%%%%%%%%%%%%%%%%%%%%%%

%%%%%%%%%%%%%%%%%%%%%%%%%%%%%%%%%%%%%%%%%%%%%%%%%%%%%%%%%%%%%%%%%%%%%%%%
\begin{figure}[t] %  figure placement: here, top, bottom, or page
   \centering
			\includegraphics[width=3in]{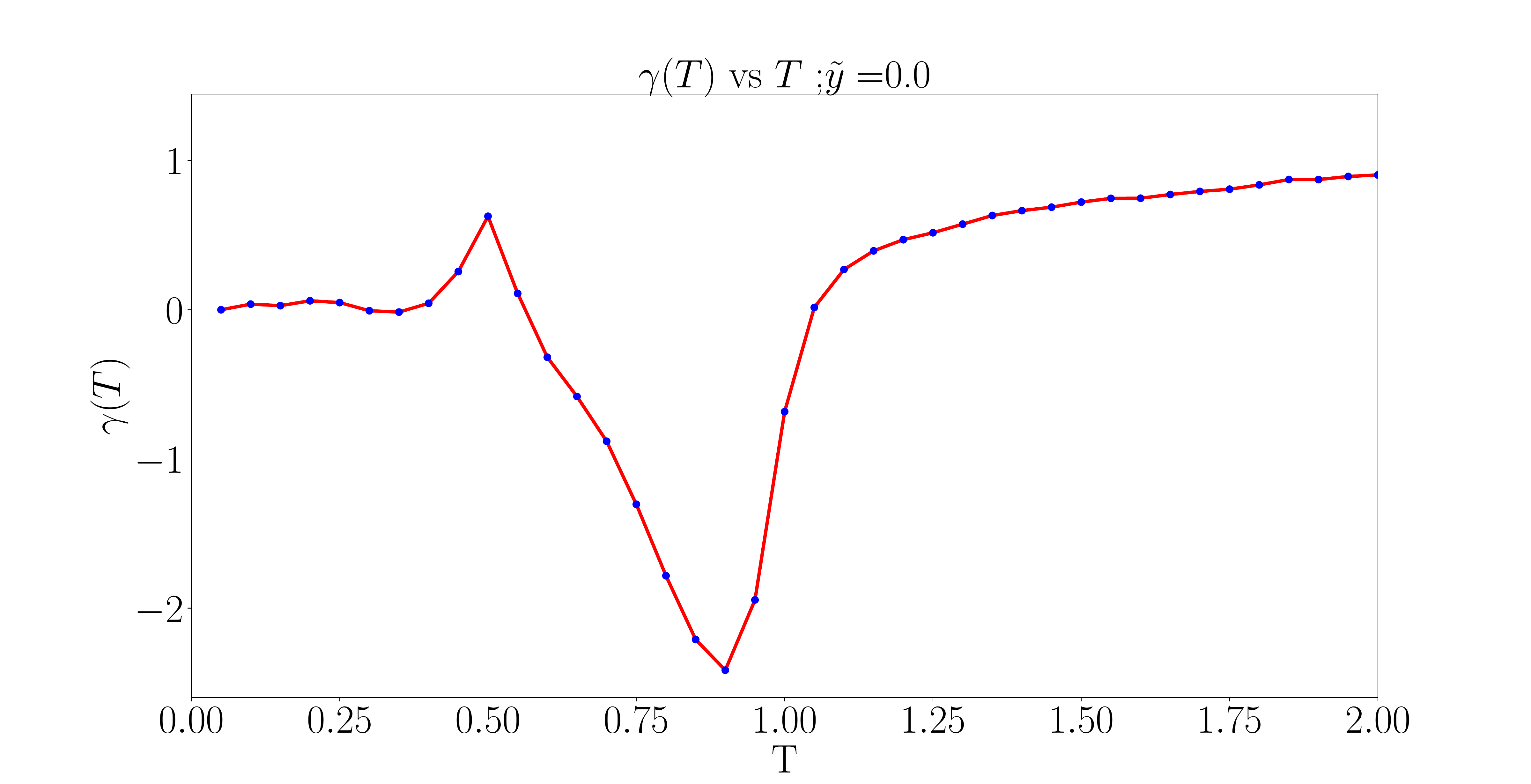} 
			\includegraphics[width=3in]{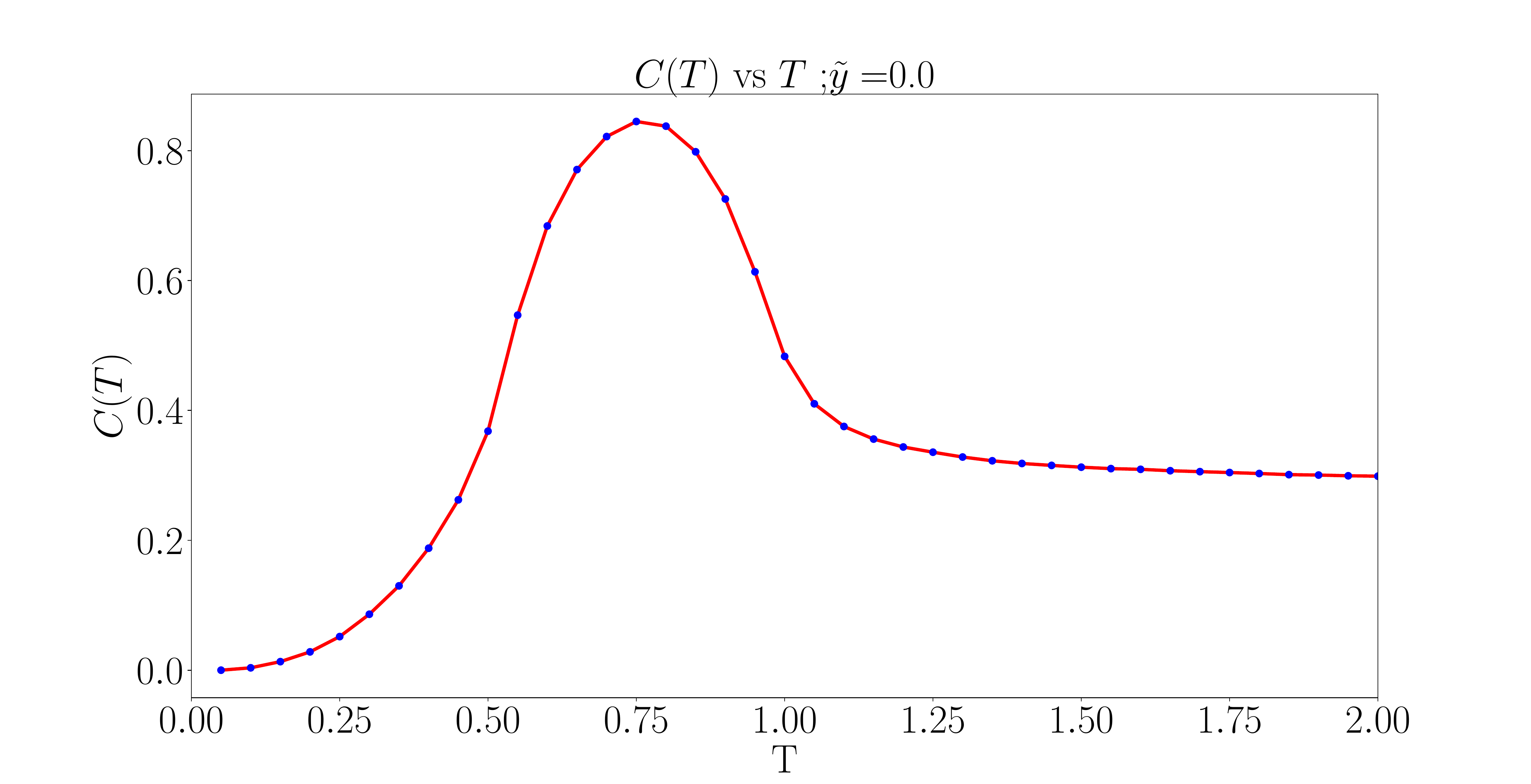} 
   \caption{The fitting of $I(X;Y,T)$ to $C(T)N+\gamma(T)$ for $\tilde y=0$. The data is obtained from fitting lattice sizes $N=8$ to $N=56$.}
	\label{C and gamma for y 0}
\end{figure}
%%%%%%%%%%%%%%%%%%%%%%%%%%%%%%%%%%%%%%%%%%%%%%%%%%%%%%%%%%%%%%%%%%%%%%%%%%%%	

Next, we further examine the case $\tilde y=0$ for different lattice sizes. The results are shown in Figure \ref{finite size scaling y is 0} for $N=8$ to $N=56$. We see that all the curves collapse onto a single curve for large values of $N$. This behavior is consistent with the assertion that R\'enyi mutual information follows the area law scaling $I(X;Y;T)={\cal C}(T)\ell+\gamma(T)$, where ${\cal C}(T)$ and $\gamma(T)$ are temperature-dependent coefficients. This behavior holds even at criticality and can be used to extract the critical temperature, as we will see momentarily.  As we discussed at the end of Section \ref{The T-dual lattice description}, the replicated system exhibits two critical temperatures at $T_c/2$ and $T_c$. The system becomes scale invariant at these two points. Therefore, we expect $I(X;Y,T)/\ell$ to be a constant for all lattice sizes, and hence, $\gamma(T)$ is expected to cross zero near $T_c/2$ and $T_c$. This behavior is easily seen in Figure \ref{C and gamma for y 0}, where we fit ${\cal C}(T)$ and $\gamma(T)$ using RMI data from $N=16$ to $N=56$. We also see that ${\cal C}(T)$ attains the asymptotic shape of Figure \ref{finite size scaling y is 0}. This explains the crossing of RMI curves and then their fan out at $T_c/2$ and $T_c$. Thus, the finite size scaling of RMI  can be used as a probe to search for phase transitions \cite{PhysRevB.87.195134}. Interestingly enough, the model given by (\ref{lattice dual XY general}) and $\tilde y=0$ (the T-dual XY spin model with no $\mathbb Z_p$ symmetry-preserving perturbation) does not have an order parameter that can be used to study the BKT phase transition\footnote{This is true despite the fact that the Hamiltonian of the system is invariant under a global $U(1)$ symmetry. The absence of symmetry breaking in XY model, or its T-dual description, is a result of the Mermin-Wagner theorem, which prohibits continuous symmetry breaking in $D\leq 2$.}. Instead, one traditionally uses the spin stiffness to accurately estimate the critical temperature \cite{Nelson:1977zz,0305-4470-38-26-003}. RMI provides an alternative probe to accurately study phase transitions in this mode, see \cite{PhysRevB.87.195134} for more details. We elaborate more on this point below.    
  
%%%%%%%%%%%%%%%%%%%%%%%%%%%%%%%%%%%%%%%%%%%%%%%%%%%%%%%%%%%%%%%%%%%%%%%%
\begin{figure}[t] %  figure placement: here, top, bottom, or page
   \centering
			\includegraphics[width=5in]{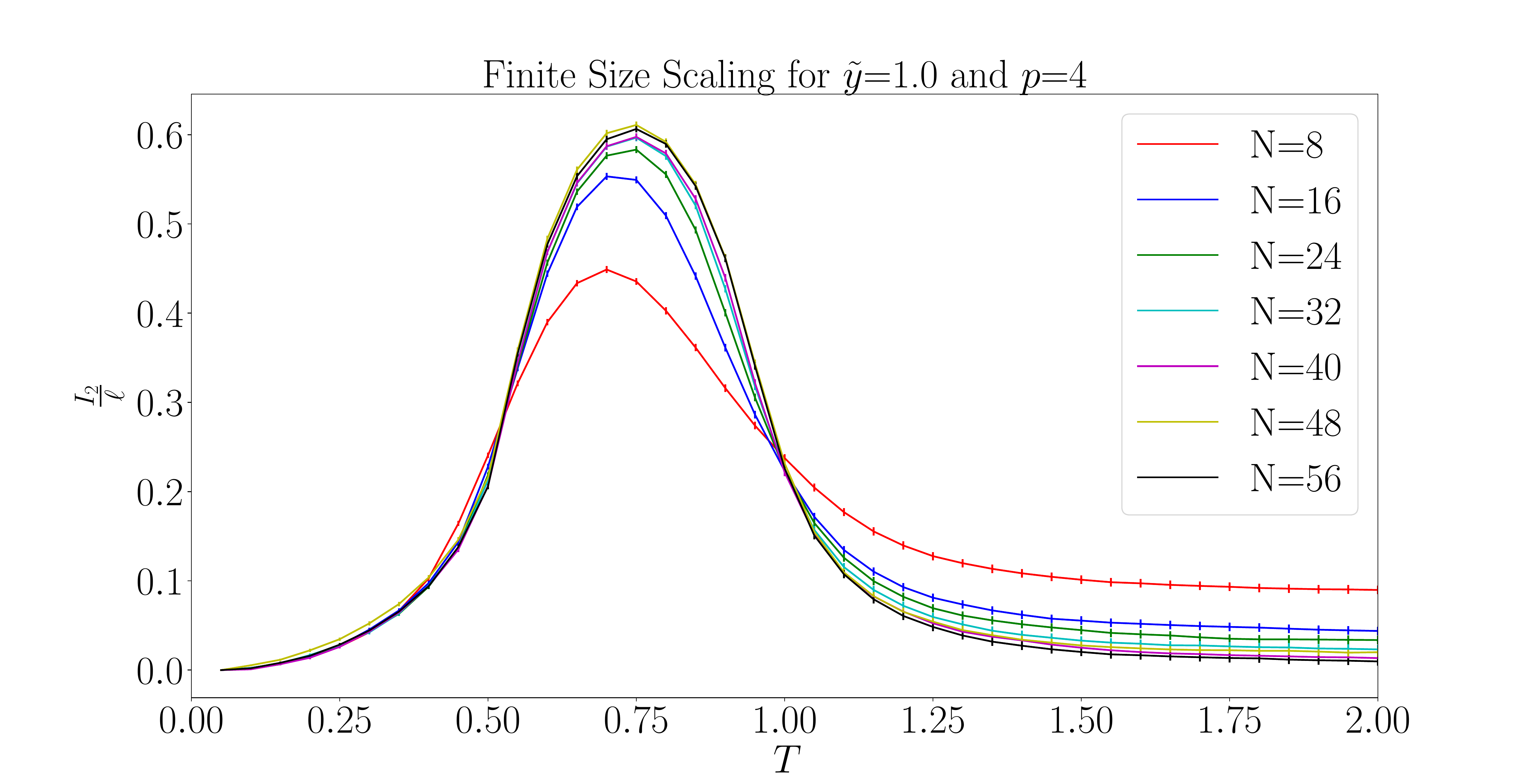} 
			\includegraphics[width=5in]{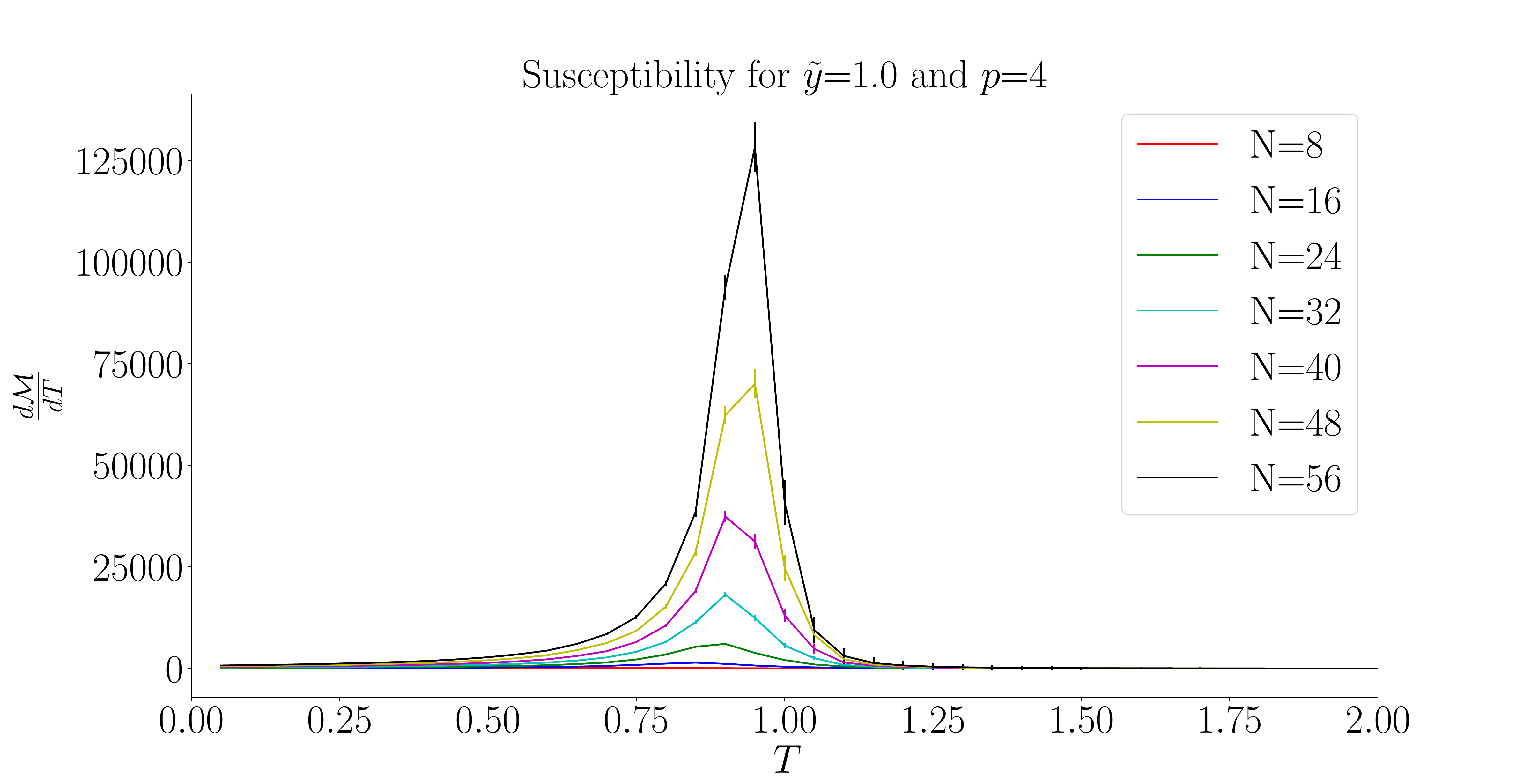} 
   \caption{Top panel: the finite size scaling for RMI of (\ref{lattice dual XY general}) with $p=4$ and $\tilde y=1$. This case corresponds to QCD(adj). The curves cross at $T\cong 0.5$ and $T\cong1$, which are the values of $T_c/2$ and $T_c$, respectively. Bottom panel: the magnetic susceptibility of the system. The susceptibility peaks at $T\cong 1$, in agreement with RMI.}
	\label{finite size scaling y qcdadj}
\end{figure}
%%%%%%%%%%%%%%%%%%%%%%%%%%%%%%%%%%%%%%%%%%%%%%%%%%%%%%%%%%%%%%%%%%%%%%%%%%%%	

%%%%%%%%%%%%%%%%%%%%%%%%%%%%
\subsection*{QCD(adj)}
%%%%%%%%%%%%%%%%%%%%%%%%%%%%

Now, we move to the finite size scaling of QCD(adj). RMI for different lattice sizes of this theory is depicted in the top panel of Figure \ref{finite size scaling y qcdadj}, where we used $\tilde y=1$ for our study. The curves cross at $T_c/2$ and $T_c$ with $T_c\cong 1$. We also calculate the magnetic susceptibility of the system, which is given by
\begin{eqnarray}
\chi_M=\frac{d|M|}{dT}\,,\quad M=\sum_{j}^{N^2}e^{i\theta_j}\,.
\end{eqnarray}
QCD(adj) is invariant under $\mathbb Z_4$ symmetry: $\theta_j\rightarrow \theta_j+\frac{2\pi}{4}$, while $M \rightarrow iM$ under the same symmetry.  Therefore, $|M|$ and $\chi_M$ are good order parameters of the system. We plot $\chi_M$ in the bottom panel of Figure \ref{finite size scaling y qcdadj}. We see that the susceptibility peaks at $T_c\cong 1$, in agreement with RMI calculations. Comparing Figures \ref{finite size scaling y is 0} and \ref{finite size scaling y qcdadj}, we see that the transition temperature is independent of $\tilde y$. This is in disagreement with the calculations of the transition temperatures in Section (\ref{The dual Sine-Gordon model and deconfinement}). One can see from the discussion of the dual Sine-Gordon model and Figures (\ref{figure EE}) and (\ref{figure EE for QCDadj}) that $T_{c\,,y=0}=2T_{c\,,QCD(adj)}$. This disagreement, however, should not come as a surprise since unlike the dual Sine-Gordon model, where both electric and magnetic fugacities are explicit parameters, the magnetic core energy of (\ref{lattice dual XY general}) is not an under-control explicit parameter. In fact, the transition temperature of the XY-spin models have only a mild dependence on the electric fugacity, as was also found in previous studies \cite{PhysRevB.69.174407}. 

\begin{figure}[t] %  figure placement: here, top, bottom, or page
   \centering
			\includegraphics[width=3in]{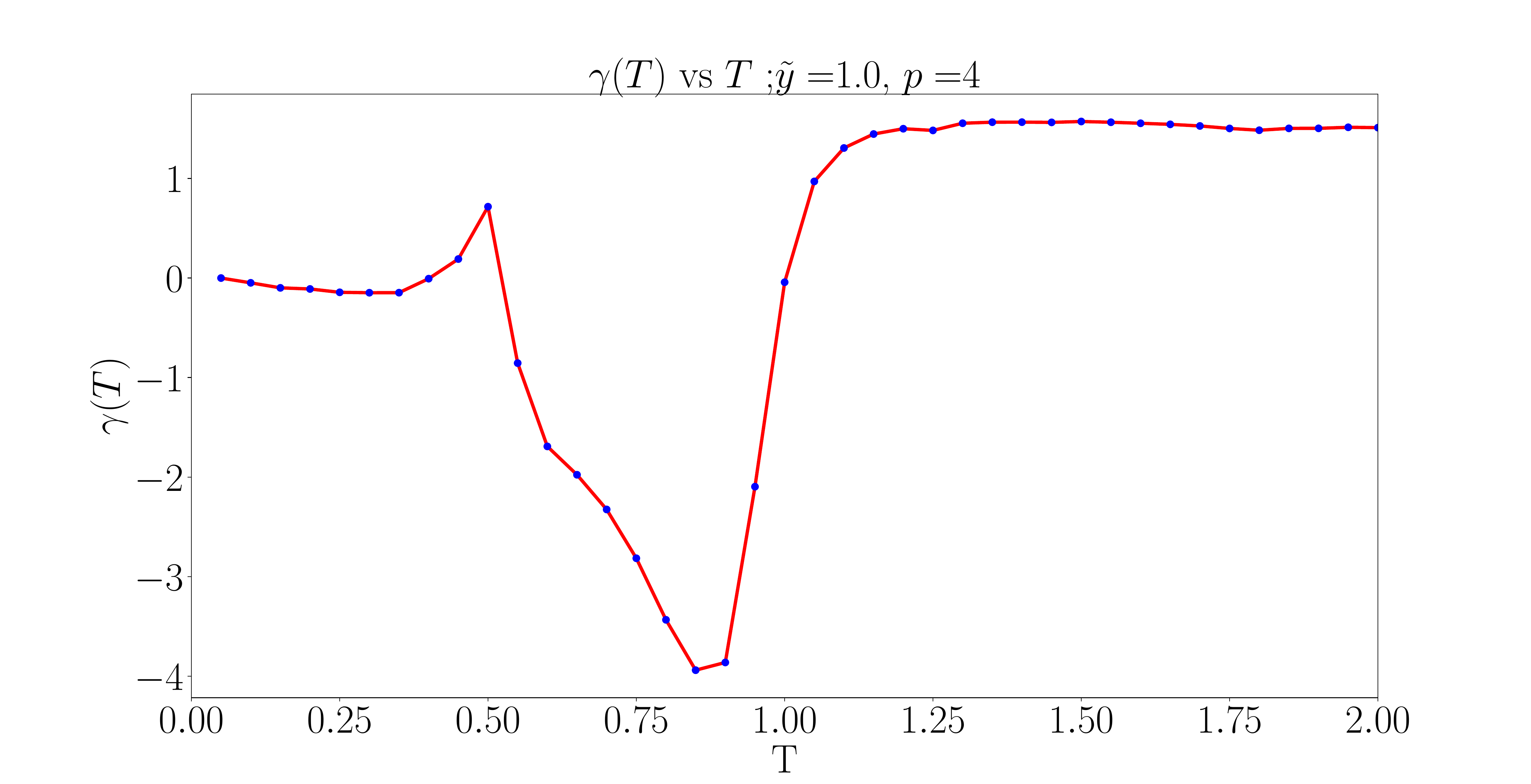} 
			\includegraphics[width=3in]{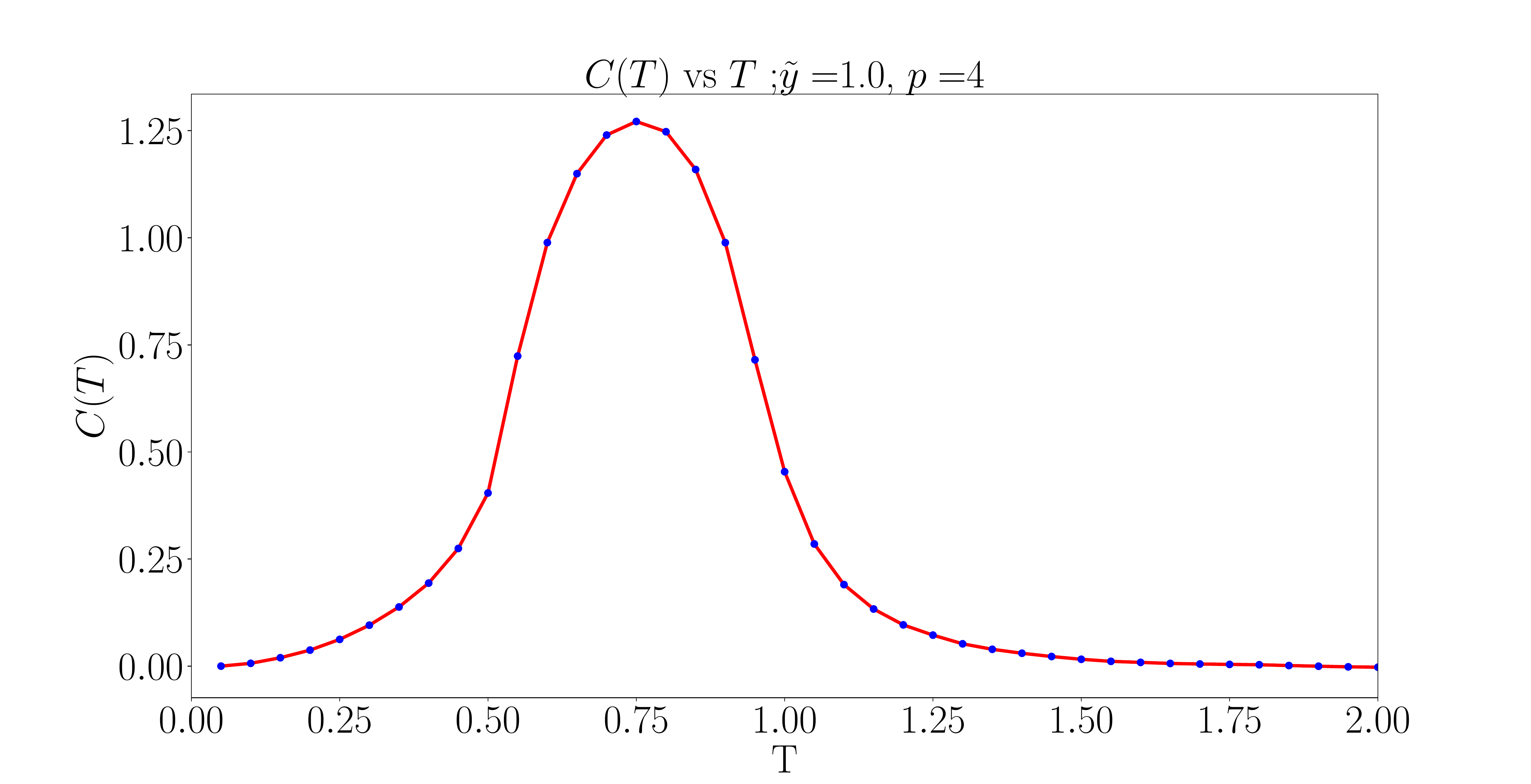}  
   \caption{The fitting of $I(X;Y,T)$ to $C(T)N+\gamma(T)$ for $p=4$ and $\tilde y=1$. The data is obtained from fitting lattice sizes $N=8$ to $N=56$.}
	\label{C and gamma}
\end{figure}

We also fit $I(X;Y;T)$ to the form $C(T)N+\gamma(T)$. The results are shown in Figure \ref{C and gamma}. It is clear that $\gamma(T)$ changes signs at $T_c/2$ and $T_c$, while $C(T)$ attains the asymptotic shape of RMI in Figure \ref{finite size scaling y qcdadj}. This explains the crossing of RMI curves at these two points, similar to the case $\tilde y=0$.

%%%%%%%%%%%%%%%%%%%%%%%%%%%%%%%%%%%%%%%%%%%%%%%%%%%%%%%%%%%%%%%%%%%%%%%%
%%%%%%%%%%%%%%%%%%%%%%%%%%%%%%%%%%%%%%%%%%%%%%%%%%%%%%%%%%%%%%%%%%%%%%%%%%%%	

%%%%%%%%%%%%%%%%%%%%%%%%%%%%%%%
\subsection*{dYM and dYM(F)}
%%%%%%%%%%%%%%%%%%%%%%%%%%%%%%%

\begin{figure}[t] %  figure placement: here, top, bottom, or page
   \centering
			\includegraphics[width=5in]{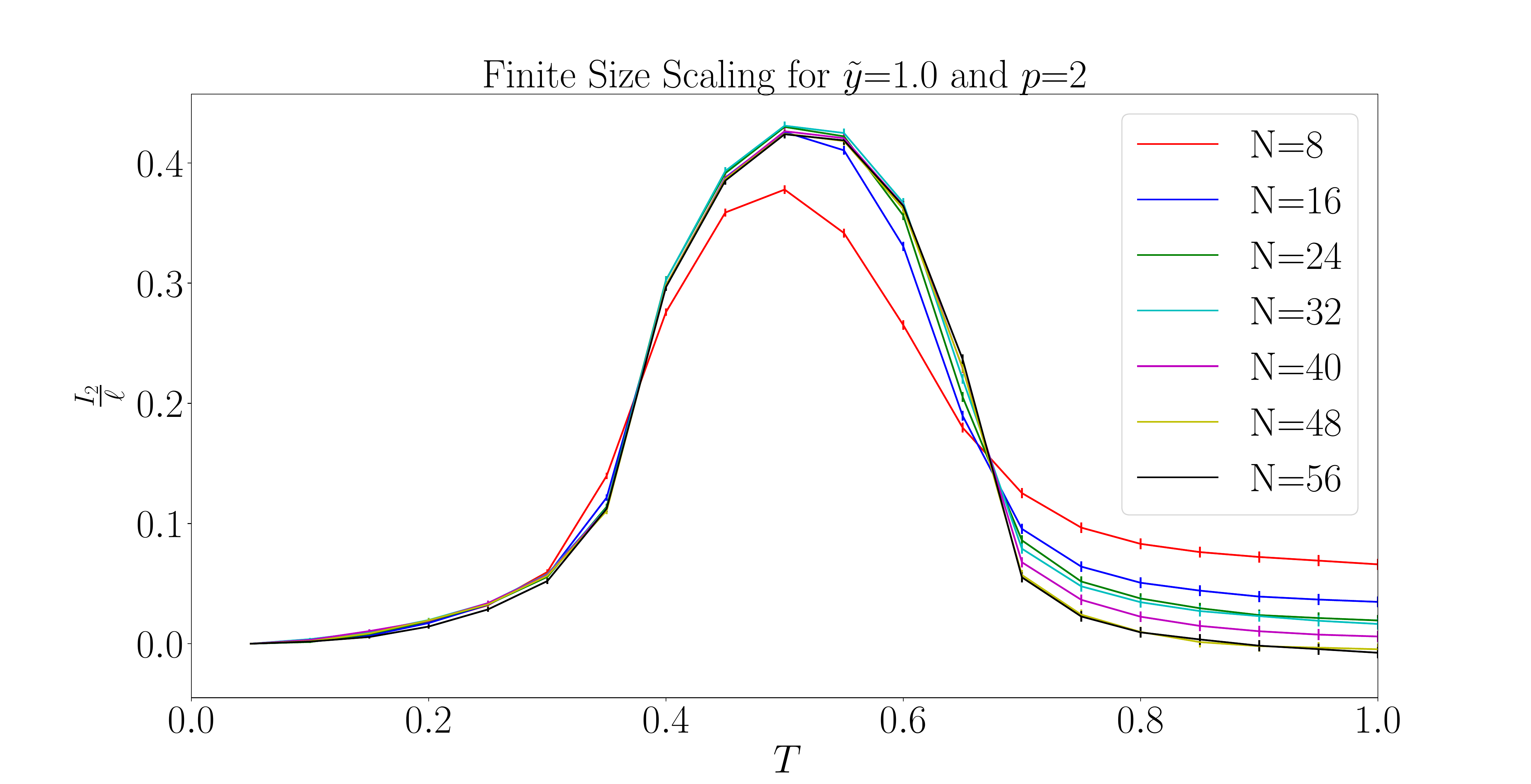} 
			\includegraphics[width=5in]{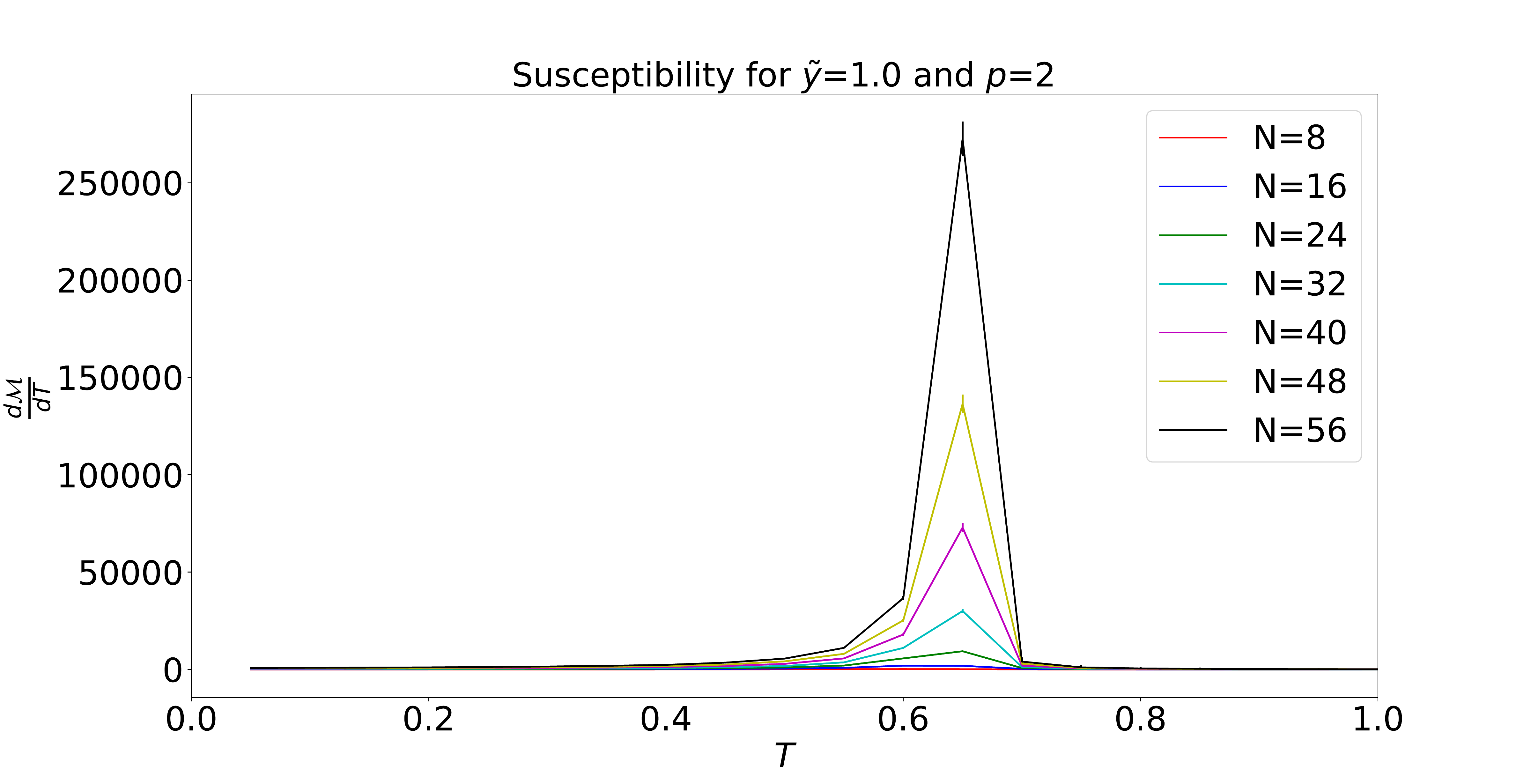}
   \caption{Top panel: the finite size scaling for RMI of (\ref{lattice dual XY general}) with $p=2$ and $\tilde y=1$. This case corresponds to dYM. The curves cross at $T\cong 0.32$ and $T\cong 0.65$, which are the values of $T_c/2$ and $T_c$, respectively. Bottom panel: the magnetic susceptibility of the system. The susceptibility peaks at $T\cong 0.65$, in agreement with RMI.}
	\label{RMI for dYM}
\end{figure}

We repeat the above analysis for dYM, $p=2$, and dYM(F), $p=1$. dYM is invariant under $\mathbb Z_2$ symmetry: $\theta_j\rightarrow \theta_j+\frac{2\pi}{2}$ and $|M|$ and $\chi_M$ are good order parameters of the system. RMI and magnetic susceptibility of dYM with $\tilde y=1$ are shown in Figure \ref{RMI for dYM} . Again, the peak of the susceptibility coincides with the second crossing of RMI curves indicating that the latter can probe phase transitions in this system. 

\begin{figure}[t] %  figure placement: here, top, bottom, or page
   \centering
			\includegraphics[width=3.5in]{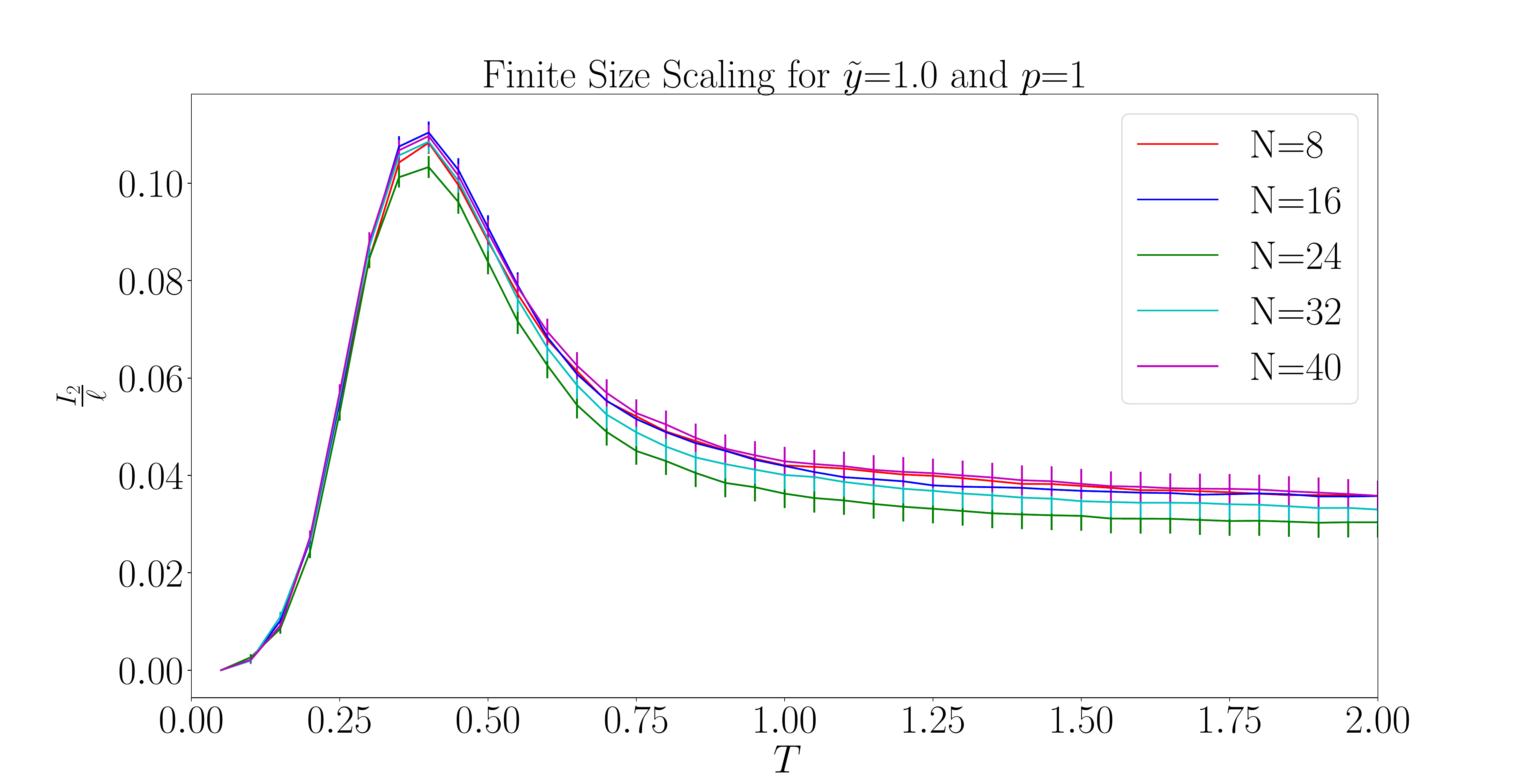}
	   \caption{RMI for $p=1$ and $\tilde y=1$. This case corresponds to dYM(F). Unlike the previous cases, RMI of different sizes does not show any features of a phase transition.}
	\label{RMI for dYMF}
\end{figure}

On the other hand, dYM(F) does not entertain any global symmetry.  RMI of dYM(F) with $\tilde y=1$ is shown in Figure \ref{RMI for dYMF}. Unlike all previous cases, RMI of different lattice sizes do not show any features of a phase transition. Also, the amplitude of RMI for $p=1$ is suppressed compared to that of $p>1$. We anticipate that this behavior is tied to the absence of global symmetries in this theory and that the theory experiences a smooth crossover from one phase to the other.   We further comment on this behavior in the Discussion Section. 
   
%%%%%%%%%%%%%%%%%%%%%%%%%%%%%%%%%%%%%%%%%%%%%%%%%%%%%%%%%%%%%%%%%%%%%%%%
%%%%%%%%%%%%%%%%%%%%%%%%%%%%%%%%%%%%%%%%%%%%%%%%%%%%%%%%%%%%%%%%%%%%%%%%%%%%

%%%%%%%%%%%%%%%%%%%%%%%%%%%%%%%%%%%%%%%%%%%%%%%%%%%%%%%%%%%%%%%%%%%%%%%%
%%%%%%%%%%%%%%%%%%%%%%%%%%%%%%%%%%%%%%%%%%%%%%%%%%%%%%%%%%%%%%%%%%%%%%%%%%%%

%%%%%%%%%%%%%%%%%%%%%%%%%%%%%%%%%%%%%%%%%%%%%%%
\section{Discussion and future directions}
\label{Discussion and future directions}
%%%%%%%%%%%%%%%%%%%%%%%%%%%%%%%%%%%%%%%%%%%%%%%

In this paper we studied the deconfinement transition in Yang-Mills theory on $\mathbb R^2\times \mathbb T^2$ by means of information-theoretic techniques in the continuum and on the lattice. The entanglement entropy calculations were achieved in the continuum via mapping the theory to a dual Sine-Gordon model. We found that this quantity attains a maximum value in  dYM, dYM(F), and QCD(adj) at the transition/crossover point. The maximum is attributed to the interchange of the role of both the magnetic and electric charges.    We also calculated R\'enyi mutual information (RMI) using a lattice version of the XY-spin model with ${\mathbb Z_p}$ symmetry-preserving perturbations. Unlike the entanglement entropy, which only captures the amount of uncertainty about the system, mutual information gives a quantitative measure of the information shared between different parts of the system.   Our RMI study is free from ambiguities that usually plague lattice gauge theories due to the non factorizability of the gauge invariant Hilbert space. We found that RMI follows the area law scaling, with subleading corrections, and their finite size scaling can be used to search for phase transitions in our theories. In particular, there is a clear crossing of RMI curves at the transition temperature in both dYM and QCD(adj), while the addition of fundamental matter washes out the crossing and dilute the information that can be shared between the system parts. As a byproduct, we also found a new method to efficiently extract RMI without the need to suppress low-winding vortices. This is done by using a T-dual description of the XY-spin model. The web of dualities in our work is tied up to the fact that Yang-Mills theory on $\mathbb R^2\times \mathbb T^2$ (with deformations or adjoint fermions) can be mapped to a dual Coulomb gas, which faithfully captures all the effective degrees of freedom near the deconfinement transition. 

dYM has a $\mathbb Z_2^C$ center symmetry that breaks in the deconfined phase. On the other hand, QCD(adj) enjoys  $\mathbb Z_2^C$ center and $\mathbb Z_2^{d\chi}$ discrete chiral symmetries. $\mathbb Z_2^{d\chi}$ is broken in the low temperature phase and gets restored in the deconfined phase. The renormalization group calculations conducted in \cite{Anber:2011gn} and our simulations indicate that the breaking of $\mathbb Z_2^C$ and restoration of $\mathbb Z_2^{d\chi}$ occurs at exactly the same critical temperature. In fact there is a constraint on the order of the occurrence of deconfinement and discrete chiral symmetry restoration in gauge theories: $T_{\scriptsize \mbox{decon}}\leq T_{\scriptsize\mbox{chiral}}$. This inequality is implied from the $\mathbb Z_2^{d\chi}$-$\left[\mathbb Z_2^C\right]^2$  mixed 't Hooft anomaly, as was shown in \cite{Shimizu:2017asf,Komargodski:2017smk}. Deconfinemnet in QCD(adj) on $\mathbb R^2 \times \mathbb T^2$ saturates this inequality. 

Adding fundamental matter to dYM breaks the center explicitly.  In this case RMI does not reveal any feature near the crossover, which is otherwise captured by the entanglement entropy of the dual Sine-Gordon model. We also found that RMI of pure XY-spin model (no $\mathbb Z_p$ symmetry-preserving perturbations) captures the transition, while entanglement entropy doesn't show any specific feature near the transition.

Before concluding our work, it is amusing to reflect on the role RMI could have played in 2-D physics had we learned about it  half a century ago. First, let us note that Mermin-Wagner theorem was published in 1966, 6 years before the discovery of the BKT phase transition. This theorem forbids continuous phase transitions in 2D, and hence, BKT phase transition in XY model came as a surprise to the physics community in 1973. Had people calculated RMI (which was not known by that time, at least among the physics community) of XY model with different $Z_p$ symmetry-preserving perturbations before 1973, they would have revealed that pure XY model (with no perturbations) is in tension with Mermin-Wagner theorem. For any $p\neq 1$ there is a discrete symmetry and symmetry breaking can happen (Mermin-Wagner theorem is no-go only for continuous symmetries). The crossing of RMI curves at a certain temperature signals the breaking of the $\mathbb Z_p$ symmetry. When $p=1$, on the other hand, the system doesn't enjoy any kind of symmetry, and hence, no crossing of RMI should be expected. This is exactly what we see in our simulations. The striking thing, however, is when we set the perturbations to zero.  Although there is a $U(1)$ symmetry in one of the phases, Mermin-Wagner theorem forbids genuine symmetry breaking. RMI curves, on the other hand, have  a clear crossing indicating that there is a nontrivial transition in the system. This is what Berezinskii, Kosterlitz, and Thouless discovered in 1973.

On the gauge theory side, we know on symmetry grounds that the presence of fundamental quarks eliminates the possibility of using the Polyakov's loop as a probe to detect phase transformations. One, however, could argue that near the transition (or crossover) the confined pairs of fundamentals and W-bosons would simply liberate making no striking difference between the presence and absence of fundamentals in the picture. Contrary to this expectation,  our simulations indicate that the presence of fundamentals makes a dramatic difference, at least from information theory point of view. This points to a tantalizing link between the absence/presence of symmetries and information stored in a system.

\subsection*{Future directions}

\begin{enumerate}
\item The special case of QCD(adj) on $\mathbb R^3\times \mathbb T^2$ with a single fermionic flavor is ${\cal N}=1$ supersymmetric glue dynamics. Deconfinement in this theory  was extensively discussed in \cite{Anber:2013doa} with conclusions similar to that of QCD(adj). The computation of RMI in this theory near the critical temperature will be discussed in a future work. We expect, however, that supersymmetry will not greatly affect the conclusions of our present work.  
\item In \cite{Anber:2012ig} an $SU(3)$ QCD(adj) theory on $\mathbb R^2 \times \mathbb T^2$ was studied via the dual Coulomb gas/ XY-spin model duality, and it was concluded that the deconfinement transition is first order. It will be interesting to examine whether RMI can have a nontrivial behavior at the transition point in this system. 
\item Our work has also applications beyond gauge theory. The study of RMI to identify classical transitions was first applied to the Ising and XY-models in \cite{PhysRevB.87.195134} and later extended to other systems like the classical toric code model \cite{PhysRevB.92.125144}. In fact, XY-spin models with perturbations are universal models that have a wide range of applications from the roughing transitions to the 2-D solid melting, see \cite{Wen:2004ym}. The calculations of RMI may help in identifying interesting features near the phase transition in these systems.  
\item Another interesting quantity that can be readily measured in XY-spin systems is the topological entanglement entropy, the constant term in $S=C\ell+\gamma_{\scriptsize \mbox{top}}$. This quantity is nonzero in systems that exhibit topological order, and hence, can be descried by topological field theories (TFT) deep in the IR. The existence of discrete 't Hooft anomalies in QCD(adj) suggests that this theory may admit a TFT that saturates the anomaly. The topological entanglement entropy can be calculated using either Levin and Wen \cite{PhysRevLett.96.110405} or Kitaev and Preskil \cite{PhysRevLett.96.110404} schemes.  Whether $\gamma_{\scriptsize \mbox{top}}$ is non-vanishing in QCD(adj) is left for a future investigation. 
\item Information about the CFT universality class can be extracted from RMI at the critical temperature. This can be done by computing RMI for different partitions of a given lattice size and trying to fit the next to leading term of RMI (the leading term is being the area) with general known behavior of CFT at finite interval. This will elucidate the link between the entanglement entropy of the dual Sine-Gordon model (which we examined in this work via CFT with deformations) and RMI of the XY models. 
\item Finally, it will be interesting to compute RMI of the full scale 4-D theory on the lattice and examine whether this quantity has a similar behavior, near the transition, to the one found in this work. 
\end{enumerate}

%%%%%%%%%%%%%%%%%%%%%%%%
\acknowledgments
%%%%%%%%%%%%%%%%%%%%%%%%

M.A. thanks Pavel Buividovich and Erich Poppitz for useful discussions and critical comments on the manuscript. We also thank William Garrick for help with Coeus cluster. This research is supported by NSF grant PHY-1720135 and the Murdock Charitable Trust.  This work was made possible in part thanks to Portland Institute for Computational Science and its resources acquired using NSF Grant \# DMS 1624776 and ARO Grant \#W911NF-16-1-0307.

\bibliography{References}

\bibliographystyle{JHEP}

\end{document}